\documentclass{aa}
\usepackage{graphicx}
\usepackage{amsmath,amssymb,amsfonts} 
\usepackage{color}                   
\usepackage{epsfig}
\usepackage{amsbsy} 

\newcommand{\Msun}{M_\odot}

\newcommand{\bu}{\boldsymbol{u}}

\newcommand{\bomega}{\boldsymbol{\omega}}
\newcommand{\bdot}{\boldsymbol{\cdot}}
\newcommand{\bnabla}{\boldsymbol{\nabla}}

\def\b{\beta}
\def\cs{c_{\rm s}}

\begin{document}

\title{Dynamical Friction in a Gas: The Supersonic Case}

\author{Aaron T. Lee\inst{\ref{inst1}} \and Steven W. Stahler\inst{\ref{inst1}}}

\institute{1 - Astronomy Department, University of California,
Berkeley, CA 94720\label{inst1}}


\date{submitted: 17 October 2013 / accepted: 8 November 2013 }

\abstract{
Any gravitating mass traversing a relatively sparse gas experiences a retarding
force created by its disturbance of the surrounding medium. In a previous
contribution \citep{ls11}, we determined this dynamical friction force when the
object's velocity was subsonic. We now extend our analysis to the supersonic
regime. As before, we consider small perturbations created in the gas far from
the gravitating object, and thereby obtain the net influx of linear momentum 
over a large, bounding surface. Various terms in the perturbation series 
formally diverge, necessitating an approximate treatment of the flow 
streamlines. Nevertheless, we are able to derive exactly the force itself. As 
in the subsonic case, we find that \hbox{$F\,=\,{\dot M}\,V$}, where $\dot M$ 
is the rate of mass accretion onto the object and $V$ its instantaneous 
velocity
with respect to distant background gas.
Our force law holds even when the object
is porous (e.g., a galaxy)
or is 
actually expelling mass in a wind. Quantitatively, the force in the supersonic
regime is less than that derived analytically by previous researchers, and is 
also less than was found in numerical simulations through the mid 1990s. We 
urge simulators to revisit the problem using modern numerical techniques. 
Assuming our result to be correct, it is applicable to many fields of 
astrophysics, ranging from exoplanet studies to galactic dynamics.
}


\titlerunning{Dynamical Friction in a Gas}
\authorrunning{Lee \& Stahler}

\maketitle

\section{Introduction}

Whenever a massive object passes through a rarefied medium, it draws surrounding
matter toward it. As a result, this material creates an overdense wake behind 
the object that exerts its own gravitational pull, retarding the original 
motion. Such dynamical friction arises whether the medium consists of 
non-interacting point particles, e.g., a stellar cluster, or a continuum 
fluid, e.g., an interstellar cloud. \citet{c43} provided the essential theory 
when the background is collisionless, and his solution has been extensively 
used in studies of both star clusters and galaxies. Interaction of a 
gravitating object with gas also occurs in a wide variety of situations. A 
partial list of topics and references includes: the interaction of planets and
gaseous disks \citep{ttp13}; the orbital decay of common-envelope binaries 
\citep{rt08}; the settling of massive stars in dense molecular clouds 
\citep{c10}; the coalescence of massive black holes in both isolated galactic 
nuclei \citep{n00} and colliding galaxies \citep{an05}; and the heating of 
intracluster gas by infalling galaxies \citep{e04}.

Despite the widespread occurrence of gaseous dynamical friction, there is still
no generally accepted derivation of the force, even after 70 years of 
effort. The flow in the vicinity of the gravitating mass is complex both 
temporally and spatially, as many simulations have shown 
\citep[see, e.g.][]{m87}. Theorists seeking a fully analytic expression for the
time-averaged force have employed various strategems to circumvent a detailed 
description of this region. In the course of their classic studies of stellar 
accretion, \citet{bh44} took the star to be traversing a zero-temperature gas. 
In their model, fluid elements follow hyperbolic orbits in the star's reference
frame and land in an infinitely thin, dense spindle behind the object. 
Bondi \& Hoyle analyzed the transfer of linear momentum from the spindle to the
star, and hence obtained the force. \citet{dm52} similarly investigated this 
hypersonic limit, as did, much more recently, \citet{c11}. \citet{d64} first 
treated a finite-temperature gas. He determined the force by integrating the 
total power emitted by the object in acoustic waves  \citep[see also][]{rs80}.
\citet{rs71} used an impulse approximation in the reference frame of the 
background gas. Finally, \citet{o99} calculated the force by integrating 
directly over the wake, whose density she obtained through a linear 
perturbation analysis.

These researchers focused principally on the supersonic case, which is often the
most interesting one astrophysically. That is, they took \hbox{$V\,>\,c_s$}, \
where $V$ is the speed of the gravitating mass relative to the distant 
background, and $c_s$ the sound speed in that gas. (In the hypersonic 
calculations of Bondi, Hoyle, and their successors, $c_s$ was implicitly set to
zero.) While their answers differed in detail, all agreed that the friction 
force in this regime varies as $V^{-2}$, with a coefficient that includes a 
Coulomb logarithm. This latter term also appears in the force derived 
by \citet{c43}, and arises from an integration in radius away from the mass. In
all derivations, at least one of the integration limits is rather ill-defined.

The previous studies made two important, simplifying assumptions. First, they
neglected any accretion of background gas by the moving object. Quantitatively,
the assumption was that \hbox{$R\,\gg\,r_{\rm acc}$}, where $R$ is the 
object's physical radius and the accretion radius 
\hbox{$r_{\rm acc}\,\equiv\,2\,G\,M/V^2$} is the distance from the mass $M$ 
within which its gravity qualitatively alters the background flow. They further
ignored $R$ compared to the scale for spatial variations in the surrounding 
flow. If $L$ denotes the latter scale, then these assumptions may be summarized
as \hbox{$L\,>>\,R\,>>\,r_{\rm acc}$}. Unfortunately, the inequality 
\hbox{$R\,>>\,r_{\rm acc}$} is often 
not
satisfied. For example, a 
low-mass star moving through a cluster-producing molecular cloud has
\hbox{$R\,\sim\,10^{11}\,\,{\rm cm}$} and 
\hbox{$r_{\rm acc}\,\sim\,10^{16}\,\,{\rm cm}$}. An analysis that covers this
regime should treat the object as being point-like in all respects, allowing 
the possibility of mass accretion through infall.

This infall cannot occur via direct impact, since the geometrical cross section 
of the body is negligible by assumption. Instead, some fluid elements that
initially miss the object are pulled back into it. Mass accretion is, in fact,
closely related to dynamical friction. In the reference frame attached to the 
mass, the background gas flows by with a speed that approaches $V$ far away. 
The steady-state accretion rate onto the object is simply the net influx of 
mass through any closed surface surrounding it. Similarly, the friction force 
is the net influx of linear momentum. Much closer to the mass, this momentum
influx manifests itself as two distinct force components. One is the 
gravitational pull from the wake, as described earlier. A second component is 
the direct advection of linear momentum from any background gas that falls into
the object. Any determination of the force by an asymptotic surface integration
cannot tease apart these two contributions.

    To clarify these ideas, Figure A.1 shows graphically the mass and momentum flow
    within the extended gas cloud surrounding the central, gravitating object. Gas
    enters the region via the dotted sphere shown in the sketch. This gas then
    follows one of three paths. Most of it leaves the sphere downstream of the
    gravitating mass, as shown by the two outermost streamlines (thin solid
    curves). Only a small fraction of the gas makes its way to the deep interior.
    Some of it accretes onto the mass, first missing it and then looping back.
    Other gas that misses more widely is temporarily slowed by the gravitational 
    pull of the mass and forms an overdense wake, also sketched in the figure. In 
    steady state, the wake cannot gain net mass, so all of the gas entering it 
    also leaves, either by joining onto the mass or exiting the sphere directly. 
    Meanwhile, the wake itself tugs on the mass gravitationally. The broad arrow 
    in the figure depicts this second form of momentum input to the mass. It is 
    possible to determine the force analytically by integrating the net momentum 
    flux over the bounding sphere. Numerical simulations can obtain this same 
    force by calculating the advective and gravitational components acting 
    directly on the mass.

In a previous paper \citep[][hereafter Paper~I]{ls11}, we did the analytic surface integration. We focused on the 
subsonic
case, 
\hbox{$V\,<\,c_s$}, which traditionally has been less explored.\footnote
{\citet{rs80} argued, based on a linear perturbation analysis, that 
the subsonic force vanishes. We showed in Paper~I that extension of their
analysis into the non-linear regime gives a finite result.} Working in the 
reference frame whose origin is attached to the mass, we developed a
perturbative method to analyze the small deviations of the background gas from 
a uniform, constant-density, flow. By integrating the perturbed variables over 
a large sphere to obtain the net momentum influx, we arrived at a surprisingly 
simple result for the force: \hbox{$F\,=\,{\dot M}\,V$}. Here, $\dot M$ is the 
mass accretion rate onto the moving object. 
To evaluate this quantity from first principles, one would have to follow the trajectories of fluid elements as they accrete onto the central mass, and ensure that, following turnaround, they smoothly cross the sonic transition. This transition occurs well inside the region of validity for our calculation. Lacking a fundamental theory for $\dot{M}$, we adopted a variant of the interpolation formula of \citet{b52} and thus found the force explicitly as a function of velocity. 
%
In this subsonic regime, our expression agrees reasonably well with past simulations \citep[e.g.,][]{r96}.

Here we apply the same technique to study the supersonic case. We quickly 
encounter the technical difficulty that some perturbation variables diverge, 
when expressed as functions of the angle from the background velocity vector.
The true flow around the gravitating mass is, of course, well-defined
everywhere, and the divergences simply indicate that the adopted series
expansions fail at certain locations. Despite this mathematical inconvenience,
we are able once again to derive the dynamical friction force through spatial
integration of the linear momentum influx. The force has exactly the same 
form as previously: \hbox{$F\,=\,{\dot M}\,V$}. In the high-speed limit, we 
recover the $V^{-2}$ behavior found by others, but not the Coulomb logarithm.

In Section~2 below, we describe our solution strategy, and then formulate the 
problem using a convenient, non-dimensional scheme. Whenever material repeats 
that of Paper~I, we abbreviate its presentation as much as possible. Section~3 
analyzes the perturbed flow to first order only. In this approximation, we find
there is no mass accretion or friction force, just as in the subsonic case. We 
extend the analysis to second order in Section~4, thus accounting for mass 
accretion. Here we also describe our method for calculating the flow 
numerically, and show sample results. Section~5 presents the analytic 
derivation of the force itself, and compares our expression to past 
simulations. Those found a greater force than we derive in the supersonic 
regime; we indicate possible causes for this discrepancy. Assuming our result
to be correct, we present several representative applications. Finally,
Section~6 compares our derivation to previous ones, and indicates directions
for future work.

\section {Method of Solution}

\subsection{Physical Assumptions}

In the reference frame whose origin coincides with the mass, the background gas
has speed $V$ far from the object and a spatially uniform density, which we 
denote as $\rho_0$. We take the gas to be isothermal, with associated sound 
speed $c_s$. As depicted in Figure~\ref{fig:coordinates}, we will be working in
a spherical coordinate system \hbox{$(r,\theta)$}, and will assume that the 
gas flow is axisymmetric about the polar ($z$-) axis, which is parallel to the 
asymptotic fluid velocity.

We neglect the self-gravity of the gas, and will be analyzing perturbations
to the flow relatively far from the central mass. Specifically, our expansions
are valid for \hbox{$r\,>>\,r_s$}, where \hbox{$r_s\,\equiv\,G\,M/c_s^2$} is
the sonic radius. We assume that in the far-field region of interest, the flow
is steady-state. Of course, a steady flow cannot be established over 
arbitrarily large distances, and one must judge, in each astrophysical
situation, whether the assumption is justified. In numerical simulations, which
we later discuss by way of comparison, the flow occurs within some fixed
computational volume, at the center of which lies the mass. For the
non-relativistic case of relevance, the force of gravity propagates 
instantaneously, but alterations in the flow density and velocity take time.
In practice, a steady flow is indeed approached in many such experiments
\citep[e.g.,][]{p00}. The observed decay of transients is instructive, but 
again is only broadly suggestive of what may occur in Nature.

Figure~\ref{fig:coordinates} singles out two special angles, both of which only
occur in supersonic flow. The first is the familiar Mach angle, defined through
the relation
\begin{equation}
{\rm sin}\,\theta_M \,\equiv\,1/\beta \,\,,
\end{equation}
where \hbox{$\beta \,\equiv\, V/c_s$}. The second is the supplement of the 
first: \hbox{$\theta_M^\prime \,\equiv\,\pi\,-\,\theta_M$}. The figure of 
revolution swept out by the radius lying along \hbox{$\theta\,=\,\theta_M$}
is the Mach cone, while we dub the analogous figure swept out by 
\hbox{$\theta\,=\,\theta_M^\prime$} the ``anti-Mach cone.'' As 
Figure~\ref{fig:regions} illustrates, we denote as the ``upstream'' region that
portion of the flow bounded by the anti-Mach cone, while the ``downstream'' 
region lies within the Mach cone. Finally, the ``intermediate'' region has
\hbox{$\theta_M\,<\,\theta\,<\theta_M^\prime$}.

A significant feature of the flow surrounding a gravitating mass is the
presence of an accretion bowshock. Any fluid element that joins onto the mass
penetrates to \hbox{$r\,\lesssim\,r_{\rm acc}$}, and encounters a shock at
\hbox{$\theta\,\sim\theta_M$}. At much larger $r$, such bowshocks weaken and 
degenerate to acoustic pulses, before fading away entirely 
\citep[][Chap.~1]{zr68}. No such shock arises near $\theta_M^\prime$, even at 
relatively small distances. In summary, for 
\hbox{$r\,\gg\,r_s\,\gtrsim\,r_{\rm acc}$}, we do not expect any discontinuities 
at $\theta_M$ or $\theta_M^\prime$ in steady-state flow, either in the fluid 
variables themselves or their derivatives. We need to keep this key point in 
mind as we encounter 
apparent
discontinuities at both angles. 

\subsection{Mathematical Formulation}

As in Paper~I, we describe the flow through two dependent variables, the mass 
density $\rho (r,\theta)$ and the stream function $\psi (r,\theta)$. Individual
velocity components may be recovered from these through the relations
\begin{eqnarray}
\label{eqn:urpsi}
u_r \,&=&\, {1\over{\rho\,r^2\,{\rm sin}\,\theta}}\,\,
{{\partial\psi}\over{\partial\theta}} \\
\label{eqn:uthetapsi}
u_\theta\,&=&\,{-1\over{\rho\,r\,{\rm sin}\,\theta}}\,\,
{{\partial\psi}\over{\partial r}} \,\,,
\end{eqnarray}
which automatically ensure mass continuity. In the extreme far-field limit, the
stream function approaches that for the background, uniform flow:
\begin{eqnarray*}
\lim_{r\,\rightarrow\,\infty}
\psi \,&=& \,{{\rho_0\,V\,r^2\,{\rm sin}^2\,\theta}\over 2}  \\
&=& \, \rho_0\,c_s\,r_s^2 \,{{\beta\,\,{\rm sin}^2\,\theta}\over 2}
\left(r\over r_s\right)^2 \,\,. 
\end{eqnarray*}

Our second form of the stream function's limit suggests how to expand 
$\psi (r,\theta)$ in a perturbation series valid for finite, but large, $r$:
\begin{eqnarray}
\psi\,&=&\,\rho_0\,c_s\,r_s^2\,\left[
f_2\left(r\over r_s\right)^2 \,+\,
f_1\left(r\over r_s\right) \,\right. \nonumber\\
&&+\,f_0\,+\, \left. f_{-1}\left(r\over r_s\right)^{-1} \,+\, ...\,\right] \,\,.
\end{eqnarray}
Here, \hbox{$f_2 \,\equiv\,\beta\,{\rm sin}^2\theta/2$}, while $f_1$, $f_0$,
$f_{-1}$, etc, are as yet unknown, non-dimensional functions of $\beta$ and 
$\theta$. We write a similar perturbation expansion for the density:
\begin{eqnarray}
\rho \,&=&\, \rho_0\,\left[ 1\,+\,
g_{-1}\left({r\over r_s}\right)^{-1} \,+\,
g_{-2}\left({r\over r_s}\right)^{-2} \, \right. \nonumber \\ &&\left.
+\, g_{-3}\left({r\over r_s}\right)^{-3} \,+\,...\,\right] \,\,.
\end{eqnarray}
The quantities $g_{-1}$, $g_{-2}$, and $g_{-3}$ are again non-dimensional 
functions of $\beta$ and $\theta$, all of them unknown at this stage.\footnote{The perturbation series in equations (4) and (5) are valid for $r >> r_s$. On
    the other hand, we expect that a smooth flow, representing a modest
    perturbation of the background stream, is present beyond $r_{\rm acc}$, which is much
    less than $r_s$ in the hypersonic regime. We also expect our final expression
    for the force to be valid as long as the background gas extends beyond $r_{\rm acc}$.}

The dynamical equations will be simplified once we recast all variables into 
non-dimensional form. We let the fiducial radius, density, and speed be $r_s$, 
$\rho_0$, and $c_s$, respectively, while we normalize the stream function to 
$\rho_0\,c_s\,r_s^2$. We will not change notation, but alert the reader whenever
we revert to dimensional varables. Our fully non-dimensional perturbation 
series are
\begin{eqnarray}
\label{eqn:exppsi}
\psi \,&=&\, f_2\ r^2\,\,+\,\,f_1\ r \,\,+\,\, f_0 \,\,\, \nonumber \\ 
&& +\, f_{-1}\ r^{-1}\,\,+\,... \\
\label{eqn:exprho} 
\rho \,&=&\,  1 \,\,+\,\,g_{-1}\ r^{-1} \,\,+\,\, g_{-2}\ r^{-2}\,\, \nonumber \\ &&+\,\,
g_{-3}\ r^{-3}\,\,+\,\,... \,\,.
\end{eqnarray}

The task will be to solve for the various functions $f_i (\theta)$ and 
$g_i (\theta)$ appearing in these two series. (Henceforth, we will suppress
the $\beta$-dependence for simplicity.) As explained in Section~2.3 of 
Paper~I, the appropriate boundary conditions are
\hbox{$f_i(\pi) \,=\, f_i^\prime (\pi) \,=\,f_i^\prime (0) \,=\, 0$} for 
\hbox{$i\,=\,1, 0, -1, -2$}, etc., and \hbox{$f_i (0) \,=\, 0$} for
\hbox{$i\,=\,1,-1,-2$}, etc. 
These conditions ensure regularity of both $u_r$ and $u_\theta$, as given by
    equations (2) and (3), respectively, on both the upstream and downstream axes.
    Since $\sin \theta \rightarrow 0$ on both axes, the $r$- and $\theta$-derivatives of $\psi$ must
    also vanish. Using the expansion of $\psi$ in equation (6), we derive the
    aforementioned conditions.
There is no associated restriction on $f_0 (0)$, which is tied to the mass accretion rate onto the
central object, as shown in Section~4 below.

\section{First-Order Flow}

\subsection{Upstream and Downstream Regions}

The $r$- and $\theta$-components of Euler's equation, in non-dimensional form,
are 
\begin{eqnarray}
\label{eqn:rEuler}
u_r\,{{\partial u_r}\over{\partial r}}\,+\,{u_\theta\over r}\,
{{\partial u_r}\over{\partial\theta}}\,-\,{{u_\theta^2}\over r}\,&=&\,\nonumber\\
-{1\over\rho}\,{{\partial\rho}\over{\partial r}}\, -\,{1\over r^2} \\
\label{eqn:thetaEuler}
u_r\,{{\partial u_\theta}\over{\partial r}}\,+\,{u_\theta\over r}\,
{{\partial u_\theta}\over{\partial\theta}}\,+\,{{u_r\,u_\theta}\over r}\,&=&\,\nonumber\\
-{1\over{\rho\,r}}\,{{\partial\rho}\over{\partial\theta}}\,\,.
\end{eqnarray}
Our procedure is first to express $u_r$ and $u_\theta$ in terms of $\psi$ and 
$\rho$, using equations~(\ref{eqn:urpsi}) and (\ref{eqn:uthetapsi}).
We then expand $\psi$ and $\rho$ themselves
in their respective perturbation series and equate the coeffcients of various 
powers of $r$. To avoid cumbersome division by the series for $\rho$, we first 
multiply equations~(\ref{eqn:rEuler}) and (\ref{eqn:thetaEuler}) through by 
$\rho^3$. 

Equating the coefficients of the highest power of $r$, which is $r^{-1}$, we 
find that they are identically equal. Equating coefficients of $r^{-2}$ yields  
the 
first-order
equations. From equation~(\ref{eqn:rEuler}), we find  
\begin{eqnarray}\label{eqn:rFO}
-\beta\,f_1^{\prime\prime} \,-\, \beta\,f_1 \,+\,
\beta^2\,{\rm sin}\,\theta\,{\rm cos}\,\theta\,\,g_{-1}^\prime \,\nonumber\\+\,
\left(\beta^2\,{\rm cos}^2\,\theta\,\,-\,1\right)\,g_{-1} \,+\,1 \,=\,0 \,\,,
\end{eqnarray}
while equation~(\ref{eqn:thetaEuler}) yields
\begin{equation}\label{eqn:thetaFO}
\left(1\,-\,\beta^2\,{\rm sin}^2\,\theta\right)\,g_{-1}^\prime \,-\,
\beta^2\,{\rm sin}\,\theta\,{\rm cos}\,\theta\,\,g_{-1}\,=\,0 \,\,.
\end{equation}
These equations are identical in 
form
to equations~(18) and (19), 
respectively, of Paper~I, hereafter designated equations~(I.18) and (I.19).
However, their 
solutions
may differ. The factor 
\hbox{$(1-\beta^2\,{\rm sin}^2\,\theta)$} in equation~(\ref{eqn:thetaFO}),
which was positive in the subsonic case for any angle $\theta$, can now be 
positive, negative, or zero.  

Consider first the upstream region. For \hbox{$\theta > \theta_M^\prime$}, the 
term \hbox{$(1-\beta^2\,{\rm sin}^2\,\theta)$} is indeed positive, and we may 
recast equation~(\ref{eqn:thetaFO}) as
\begin{equation*}
{{d{\phantom \theta}}\over{d\theta}}
\left[\left(1\,-\,\beta^2\,{\rm sin}^2\,\theta\right)^{1/2}\,g_{-1}\right]
\,=\,0\,\,, 
\end{equation*}
whose solution is
\begin{equation}\label{eqn:gFOupC}
g_{-1} \,=\, {C\over{\left(1\,-\,\beta^2\,{\rm sin}^2\,\theta\right)^{1/2}}}
\,\,.
\end{equation}
Here, $C$ is independent of $\theta$, but is possibly a function of $\beta$. The
analogous result appeared in our subsonic analysis as equation~(I.20). In that 
case, we ultimately found $C$ to be unity.

Returning to the supersonic flow, we see that, for any non-zero $C$, the
function $g_{-1}$ diverges at \hbox{$\theta \,=\,\theta_M^\prime$}. This fact 
does 
not
mean that the density itself diverges at that location; there 
is no reason for it to do so. Rather, it is our series expansion of
$\rho\,(r,\theta)$ that fails in this region. The issue here is mathematical, 
rather than physical. Although the function $g_{-1} (\theta)$ diverges, the 
associated term in the perturbation expansion, equation~(\ref{eqn:exprho}),
is multiplied by $r^{-1}$. At any finite angular separation from 
$\theta_M^\prime$, the first-order correction to the background density can be 
made arbitrarily small by considering a sufficiently large $r$-value. This 
observation suggests that the divergences encountered here and elsewhere in our
analysis will vanish if we change independent variables from $r$ and $\theta$ 
to another set that mixes the two. In any case, we do not explore that 
possibility in the present paper.

Substitution of equation~(\ref{eqn:gFOupC}) for $g_{-1}$ into (\ref{eqn:rFO})
yields the governing equation for $f_1$:
\begin{equation}
f_1^{\prime\prime} \,+\, f_1 \,=\, {1\over\beta} \,-\,
{{C\,\left(1\,-\,\beta^2\right)}\over
{\beta \left(1\,-\,\beta^2\,{\rm sin}^2\,\theta\right)^{3/2}}} \,\,.
\end{equation}
This equation, being identical to (I.21), has the same general solution:
\begin{eqnarray}\label{eqn:fFO}
f_1 \,&=&\,{1\over\beta} \,-\,
{{C\,\left(1\,-\,\beta^2\,{\rm sin}^2\,\theta\right)^{1/2}}\over\beta} \,\nonumber\\ &        &
+\,D\,{\rm cos}\,\theta \,+\, E\,{\rm sin}\,\theta \,\,,
\end{eqnarray}
where $D$ and $E$ are additional constants. The vanishing of $f_1\,(\pi)$ tells
us that \hbox{$D\,=\,(1-C)/\beta$}, while \hbox{$f_1^\prime\,(\pi)\,=\,0$} 
implies that \hbox{$E\,=\,0$}. Both relations also held in the subsonic case. 

We next turn to the downstream region. Here again, the critical term 
\hbox{$(1-\beta^2\,{\rm sin}^2\,\theta)$} is positive. The solutions for 
$g_{-1}$ and $f_1$ are thus identical to the upstream solutions in 
equations~(\ref{eqn:gFOupC}) and (\ref{eqn:fFO}), if we replace $C$, $D$, 
and $E$ by new constants $C^\prime$, $D^\prime$, and $E^\prime$. Application of the
appropriate boundary conditions then tells us that \hbox{$E^\prime\,=\,0$} and
\hbox{$D^\prime\,=\,(1-C^\prime)/\beta$}.

Referring again to Figure~\ref{fig:regions}, the supersonic flow we are now 
analyzing differs from the subsonic one by the presence of the intermediate 
region, lying between the Mach and anti-Mach cones. As $\beta$ decreases to 
unity from above, this region narrows symmetrically about 
\hbox{$\theta\,=\,\pi/2$} and ultimately vanishes, as does the physical 
distinction between the supersonic and subsonic flows. Continuity thus demands 
that, while \hbox{$(C,D)$} and \hbox{$\left(C^\prime,D^\prime\right)$} could, in 
principle, be $\beta$-dependent, they are actually not. Instead, they must have
the truly constant values \hbox{$(1,0)$} found in the subsonic case.

In summary, the first-order flow in both the upstream and downstream regions is
described by 
\begin{eqnarray}
\label{eqn:fFOup}
f_1 \,&=&\, {{1\,-\,\left(1\,-\,\beta^2\,{\rm sin}^2\,\theta\right)^{1/2}}\over
\beta} \\
\label{eqn:gFOup}
g_{-1} \,&=&\, 
{1\over{\left(1\,-\,\beta^2\,{\rm sin}^2\,\theta\right)^{1/2}}} \,\,.
\end{eqnarray}
After substituting these coefficients into the series expansions for $\psi$
and $\rho$, one may calculate $u_r$ and $u_\theta$ to linear order from
equations~(\ref{eqn:urpsi}) and (\ref{eqn:uthetapsi}), respectively. At any 
fixed, $r$-value, however, the series representation for $\rho$ fails 
sufficiently close to either Mach cone, and the velocity components cannot be 
obtained in this manner.


Comparing our results thus far with the previous literature, \citet{d64}, 
\citet{rs71}, and \citet{o99} all obtained equation~(\ref{eqn:gFOupC}) for the
downstream, first-order density distribution, but with \hbox{$C\,=\,2$}. They 
also found that $g_{-1}$ vanishes for \hbox{$\theta > \theta_M$}. In the 
subsonic analysis of \citet{o99}, $g_{-1}$ is everywhere finite. Thus, the 
density distributions in her supersonic and subsonic flows do not approach one 
another in the \hbox{$\beta\,=\,1$} limit. \citet{rs71} further argued that the
apparent divergence in the downstream density perturbation at 
\hbox{$\theta\,=\,\theta_M$} signifies the presence of a bowshock. While a
bowshock certainly arises relatively close to the gravitating mass, it does not
persist into the far field, the only regime where an analysis based on small 
perturbations of the background gas is justified. 

\subsection{Intermediate Region} 

For \hbox{$\theta_M < \theta < \theta_M^\prime$}, the term 
\hbox{$(1-\beta^2\,{\rm sin}^2\,\theta)$} in equation~(\ref{eqn:thetaFO}) is
negative. We therefore rewrite this equation as
\begin{equation}
\left(\beta^2\,{\rm sin}^2\,\theta\,-\,1\right)\,g_{-1}^\prime \,+\,
\beta^2\,{\rm sin}\,\theta\,{\rm cos}\,\theta\,\,g_{-1}\,=\,0 \,\,.
\end{equation}
This is equivalent to
\begin{equation*}
{{d{\phantom \theta}}\over{d\theta}}
\left[\left(\beta^2\,{\rm sin}^2\,\theta\,-\,1\right)^{1/2}\,g_{-1}\right]
\,=\,0\,\,,
\end{equation*}
which in turn implies that
\begin{equation}
\label{eqn:gFOiC}
g_{-1} \,=\, {C\over{\left(\beta^2\,{\rm sin}^2\,\theta\,-\,1\right)^{1/2}}}
\,\,.
\end{equation}
Here we are reverting to our original notation for the integration constants, as
the previous ones have all been evaluated.

Substitution of the new expression for $g_{-1}$ into equation~(\ref{eqn:rFO})
yields
\begin{equation}
f_1^{\prime\prime} \,+\, f_1 \,=\, {1\over\beta} \,-\,
{{C\,\left(\beta^2\,-\,1\right)}\over
{\beta \left(\beta^2\,{\rm sin}^2\,\theta\,-\,1\right)^{3/2}}} \,\,.
\end{equation}
We may solve this equation using the method of variation of parameters. Adding 
the two homogeneous solutions yields the general result
\begin{eqnarray}\label{eqn:fFOiC}
f_1 \,&=&\,{1\over\beta} \,+\,
{{C\,\left(\beta^2\,{\rm sin}^2\,\theta\,-\,1\right)^{1/2}}\over\beta} \,\nonumber\\ &&+\,
D\,{\rm cos}\,\theta \,+\, E\,{\rm sin}\,\theta \,\,.
\end{eqnarray}
As in the upstream and downstream regions, the series representation of $\psi$
is well-behaved at either Mach cone, at least to linear order. Approaching
either cone from the outside, equation~(\ref{eqn:fFOup}) tells us that
\hbox{$f_1\,\rightarrow\,1/\beta$}. Equation~(\ref{eqn:fFOiC}) above implies 
that $f_1$ has the additional term 
\hbox{$D\,{\rm cos}\,\theta\,+\,E\,{\rm sin}\,\theta$} when we approach the 
cones from the intermediate region. 

To proceed, we consider the physical interpretation of the stream function. 
Equation~(\ref{eqn:urpsi}) tells us that $\psi (r,\theta)$ is the 
$\theta$-integral of the mass flux $\rho\,u_r$ over a surface of radius $r$. 
More generally, the stream function is the net rate of mass transport into any 
surface of revolution extending from \hbox{$\theta\,=\,0$} to the angle of 
interest. A discontinuity in $\psi$ at an angle $\theta$ thus represents a thin
sheet of mass being injected or ejected along the corresponding cone. If we are
to reject such a solution as unphysical, then we must demand continuity of 
$\psi$. That is, $f_1$ must again approach $1/\beta$ at the two Mach cones, and 
\hbox{$D\,=\,E\,=\,0$} in equation~(\ref{eqn:fFOiC}).   

Our mathematical description of the first-order flow in the intermediate region
is now
\begin{eqnarray}
\label{eqn:fFOiC2}
f_1 \,&=&\, {{1\,+\,C\,\left(\beta^2\,{\rm sin}^2\,\theta\,-\,1\right)^{1/2}}
\over\beta} \\
\label{eqn:gFOiC2}
g_{-1} \,&=&\, 
{C\over{\left(\beta^2\,{\rm sin}^2\,\theta\,-\,1\right)^{1/2}}} \,\,.
\end{eqnarray}
Since the intermediate region does not reach either the upstream or downstream
axis, we cannot appeal to our usual boundary conditions in order to evaluate 
$C$. We defer this issue until our analysis of the second-order flow in 
Section~4, when we will show that the requirement of mass continuity again 
settles the matter. For any $C$-value, the divergence of $g_{-1}$ at the Mach 
and anti-Mach cones indicates a breakdown of the series representation for the 
density.

\subsection{Vorticity}

A key, simplifying property of the flow is that it is irrotational. To reprise
the argument from Paper~I, we first write Euler's equation in the steady state 
as
\begin{equation}\label{eqn:VorEuler}
\bu \times \bomega \,=\, \bnabla B  \,\,,
\end{equation}
where \hbox{$\bomega \,\equiv\bnabla\times\bu$} is the vorticity and the
Bernoulli function on the righthand side is
\begin{equation}\label{eqn:Bf}
B \,\equiv\,{1\over 2}\,u^2 \,+\, 
{\rm ln}\,\rho \,-\,\frac{1}{r}
\,\,.
\end{equation}
Dotting both sides of equation~(\ref{eqn:VorEuler}) with $\bu$, we find that
\begin{equation*}
\left(\bu\,\bdot\,\bnabla\right)\,B\,=\,0\,\,,
\end{equation*}
which is the familiar statement that $B$ is constant along streamlines. 
Moreover, $B$ approaches $\beta^2/2$ at large $r$, both upstream and downstream.
(Recall that the non-dimensional density $\rho$ becomes unity in this limit.)
In these regions, therefore, $B$ has the same value on every streamline, 
and \hbox{$\bnabla B\,=\,0$}. Since $\bomega$ is orthogonal to $\bu$ in our 
poloidal flow, equation~(\ref{eqn:VorEuler}) imples that 
\hbox{$\bomega \,=\, 0$}, i.e., the flow is irrotational outside the Mach cones.

If our derived intermediate solution is correct, then irrotationality must hold
there as well, since there is no physical barrier between the intermediate 
flow and those upstream and downstream, at least in the laminar, far field. 
The only non-zero component of the vorticity is $\omega_\phi$, so the condition
of irrotationality becomes
\begin{equation}\label{eqn:Vorphi}
{{\partial u_r}\over{\partial\theta}} \,=\,
{{\partial \left(r u_\theta\right)}\over{\partial r}} \,\,.
\end{equation}
To test the validity of this relation in the intermediate region, we use 
equations~(\ref{eqn:urpsi}) and (\ref{eqn:uthetapsi}) for $u_r$ and $u_\theta$, 
respectively, and then substitute in the series expansions for $\psi$ and 
$\rho$, equations~(\ref{eqn:exppsi}) and (\ref{eqn:exprho}).

Equating the coefficient of the highest power of $r$, which is $r^0$, is 
equivalent to testing for irrotationality in the uniform, background flow. 
Specifically, we require
\begin{equation}\label{eqn:Vor0}
f_2^{\prime\prime}\ \,-\,
f_2^\prime\,\,{\rm cot}\,\theta \,=\,
-2\,f_2 \,\,.
\end{equation}
Using \hbox{$f_2\,=\,\beta\,{\rm sin}^2\,\theta/2$}, we verify that the above
equation does hold. This result is to be expected, as a uniform flow is
manifestly irrotational.  

We next equate the coefficients of $r^{-1}$, effectively testing 
irrotationality in the first-order flow. The required condition is now
\begin{equation}\label{eqn:Vor1}
f_1^{\prime\prime}\,-\,f_2^{\prime\prime}\,\,g_{-1}\,-\,f_2^\prime
\,\,g_{-1}^\prime \,=\,
\left(f_1^\prime\,-\,f_2^\prime\,\,g_{-1}\right){\rm cot}\,\theta
\,\,,
\end{equation} 
where $f_1$ and $g_{-1}$ are given by equations~(\ref{eqn:fFOiC2}) and 
(\ref{eqn:gFOiC2}), respectively. Using these functional forms, we find that 
this last equation is satisfied for any value of $C$. Thus, the first-order 
flow in the intermediate region is indeed irrotational. 

\section{Second-Order Flow}

\subsection{Dynamical Equations}

Having established the first-order flow, at least up to the constant $C$, we
consider the next higher approximation. We return to the perturbative expansion
of Euler's equation, as described at the beginning of Section~3.1. By equating
coefficients of $r^{-3}$, we derive the 
second-order
equations, which
govern the variables $f_0$ and $g_{-2}$. Here the source terms involve $f_1$ and
$g_{-1}$. Prior to substituting in the explicit solutions for $f_1$ and 
$g_{-1}$, the equations are identical to those in the subsonic problem; we
display them again for convenient reference. 

From the $r$-component of Euler's equation, we derive equation~(I.27), which is
\begin{eqnarray}\label{eqn:rSO}
&-&\beta\,f_0^{\prime\prime} \,-\, \beta\,{\rm cot}\,\theta\,f_0^\prime \,+\,\
\beta^2\,{\rm sin}\,\theta\,{\rm cos}\,\theta\,\,g_{-2}^\prime \nonumber \\ \,&+&\,
\left(2\,\beta^2\,{\rm cos}^2\,\theta\,-\,2\right)\,g_{-2}\,\nonumber\\&=&\,
{\cal A}_1 \,+\, {\cal A}_2 \,+\, {\cal A}_3 \,\,.
\end{eqnarray}
The three righthand terms are
\begin{eqnarray*}
{\cal A}_1 \,&\equiv&\, {f_1^2\over{{\rm sin}^2\,\theta}} \,-\,
{{f_1\,f_1^\prime\,{\rm cos}\,\theta}\over{{\rm sin}^3\,\theta}} \,+\,
{{\left(f_1^\prime\right)^2}\over{{\rm sin}^2\,\theta}} \\&&\,+\,
{{f_1\,f_1^{\prime\prime}}\over{{\rm sin}^2\,\theta}}
\\ 
{\cal A}_2 \,&\equiv&\, \beta\,f_1\,\,g_{-1} \,-\,
2\,\beta\,f_1^\prime\,\,g_{-1}\,{\rm cot}\,\theta \,\\&&-\,
\beta\,f_1\,\,g_{-1}^\prime\,{\rm cot}\,\theta  \,-\,
\beta\,f_1^\prime\,\,g_{-1}^\prime \,\\&&+\,
\beta\,f_1^{\prime\prime}\,\,g_{-1} 
\\
{\cal A}_3 \,&\equiv&\, 2\,g_{-1}^2  \,-\, 3\,g_{-1}
\,\,.
\end{eqnarray*}
The $\theta$-component of Euler's equation yields equation~(I.31):
\begin{eqnarray}\label{label:thetaSO}
&-&\beta\,f_0^\prime \,+\, {\cal D}\,g_{-2}^\prime \,-\,
2\,\beta^2\,{\rm sin}\,\theta\,\,{\rm cos}\,\theta\,\,g_{-2} \,\nonumber\\&=&\,
{\cal B}_1 \,+\, {\cal B}_2 \,+\, {\cal B}_3
\,\,.
\end{eqnarray}
Here, 
\begin{equation}
{\cal D}\,\equiv\,1\,-\,\beta^2\,{\rm sin}^2\,\theta \,\,,
\end{equation}
and
\begin{eqnarray*}
{\cal B}_1 \,&\equiv&\, f_1^2\,{\rm cot}\,\theta \,-\,
{{f_1\,f_1^\prime}\over{{\rm sin}^2\,\theta}} 
\\
{\cal B}_2 \,&\equiv&\, \beta\,f_1\,\,g_{-1}\,{\rm cot}\,\theta \,\,+\,\,
\beta\,f_1^\prime\,\,g_{-1} \,\,\\&&+\,\,2\,\beta\,f_1\,g_{-1}^\prime
\\
{\cal B}_3 \,&\equiv&\, -2\,g_{-1}\,\,g_{-1}^\prime
\,\,.
\end{eqnarray*}

In the upstream and downstream regions, there are no unknown constants in the
source terms, i.e., both $f_1$ and $g_{-1}$ are identical to their subsonic
counterparts. Hence, the explicit form of the second-order equations is also the
same. Referring to equations~(I.35) and (I.36), we have
\begin{eqnarray}\label{eqn:rSOu}
&-&\beta\,f_0^{\prime\prime} \,-\, \beta\,{\rm cot}\,\theta\,f_0^\prime \,+\,\
\beta^2\,{\rm sin}\,\theta\,{\rm cos}\,\theta\,\,g_{-2}^\prime\nonumber\\  \,&+&\,
\left(2\,\beta^2\,{\rm cos}^2\,\theta\,-\,2\right)\,g_{-2}\,\nonumber\\&=&\,
{1\over{\cal D}}\,-\,{3\over\sqrt{\cal D}} \,+\,
{2\over{1+\sqrt{\cal D}}}\,\,,
\end{eqnarray} 
and
\begin{eqnarray}\label{eqn:thetaSOu}
&-&\beta\,f_0^\prime \,+\, {\cal D}\,g_{-2}^\prime \,-\,
2\,\beta^2\,{\rm sin}\,\theta\,\,{\rm cos}\,\theta\,\,g_{-2}
\nonumber \\ 
&=&\,
\beta^2\,{\rm sin}\,\theta\,\,{\rm cos}\,\theta
\left[-{2\over{{\cal D}^2}} \,+\, {2\over{{\cal D}^{3/2}}}\,\right.\nonumber\\
\,&-&\left.
{1\over{\cal D}} \,+\, {1\over{\left(1+\sqrt{\cal D}\right)^2}}
\right]\,\,.
\end{eqnarray}

In the intermediate region, however, $f_1$ and $g_{-1}$ are given by the new
expressions in equations~(\ref{eqn:fFOiC2}) and (\ref{eqn:gFOiC2}),
respectively. After lengthy manipulation, we find the new source terms, and 
hence the explicit dynamical equations in this region. These equations, which 
still contain the unknown constant $C$, are 
\begin{eqnarray}\label{eqn:rSOi}
&-&\beta\,f_0^{\prime\prime} \,-\, \beta\,{\rm cot}\,\theta\,f_0^\prime +
\beta^2\,{\rm sin}\,\theta\,{\rm cos}\,\theta\,\,g_{-2}^\prime\nonumber\\  \,&+&\, 
\left(2\,\beta^2\,{\rm cos}^2\,\theta\,-\,2\right)\,g_{-2}\,\,={{2\,C\,{\cal E}^{1/2}}\over{\beta^2\,{\rm sin}^2\,\theta}} \,\nonumber \\ 
 &-&
{{3\,C}\over{{\cal E}^{1/2}}} 
+
{C^2\over{\cal E}} \,+\,
{{1 - C^2}\over{\beta^2\,{\rm sin}^2\,\theta}} \,\,,
\end{eqnarray} 
and
\begin{eqnarray}\label{eqn:thetaSOi}
&-&\beta\,f_0^\prime \,-\, {\cal E}\,g_{-2}^\prime \,-\,
2\,\beta^2\,{\rm sin}\,\theta\,\,{\rm cos}\,\theta\,\,g_{-2} \,\nonumber\\ \,&=&
\,
{{2\,C^2\,\beta^2\,{\rm sin}\,\theta\,{\rm cos}\,\theta}\over{{\cal E}^2}}
\,-\,
{{2\,C\,\beta^2\,{\rm sin}\,\theta\,{\rm cos}\,\theta}\over{\cal E}^{3/2}}  \nonumber\\
&-&
{{C^2\,\beta^2\,{\rm sin}\,\theta\,{\rm cos}\,\theta}\over{\cal E}} +{{\left(1-C^2\right){\rm cot}\,\theta}\over{\beta^2\,{\rm sin}^2\,\theta}}
\,\,
\nonumber\\ 
&+&\, C^2\,{\rm cot}\,\theta +\,{{2\,C\,{\cal E}^{1/2}\,{\rm cot}\,\theta}\over{\beta^2\,{\rm sin}^2\,\theta}}
\, .
\end{eqnarray}
Here, we have defined
\begin{equation} 
{\cal E}\,\equiv\, \beta^2\,{\rm sin}^2\,\theta\,-\,1 \,\,,
\end{equation} 
which is positive throughout this region. 

\subsection{Near-Cone Divergence}

Our plan is to integrate numerically two, coupled second-order equations in
order to determine the variables $f_0^\prime$ and $g_{-2}$ throughout the flow. 
With $f_0^\prime$ in hand, another integration will yield $f_0$ itself. Before we
embark on this program, let us consider more carefully the behavior of 
$f_0^\prime$ and $g_{-2}$. In the upstream and downstream regions, the term 
$\cal D$ vanishes as one approaches either Mach cone. Thus, the source terms in
equations~(\ref{eqn:rSOu}) and (\ref{eqn:thetaSOu}) diverge in that limit. 
Similarly, the source terms in equations~(\ref{eqn:rSOi}) and 
(\ref{eqn:thetaSOi}) diverge because of the vanishing of $\cal E$. It is
possible, then, that both $f_0^\prime$ and $g_{-2}$ also diverge at the cones.

This is indeed the case. In this section, we first establish the divergences of 
$f_0^\prime$ and $g_{-2}$ in the downstream region only. That is, we assume that 
$\theta$ approaches $\theta_M$ from below. Later in the section, we outline the
derivation for the remaining regions of the flow. We then use the full result 
to shed light on the unknown parameter $C$.

Since \hbox{$\theta\,<\,\theta_M$}, we define the positive angle
\hbox{$\alpha\,\equiv\,\theta_M - \theta$}, and assume that $f_0^\prime$ and
$g_{-2}$ take the following asymptotic forms as $\alpha$ diminishes:
\begin{eqnarray}
\label{eqn:fSOas}
f_0^\prime \,&=&\, \zeta\,\,\alpha^{-m} \\
\label{eqn:gSOas}
g_{-2} \,&=&\, \gamma\,\,\alpha^{-n} \,\,,
\end{eqnarray}
where $\zeta$, $\gamma$, $m$, and $n$ are all functions of $\beta$ only. If both
$f_0^\prime$ and $g_{-2}$ truly diverge, then $m$ and $n$ are positive. Within 
equations~(\ref{eqn:rSOu}) and (\ref{eqn:thetaSOu}), we replace 
${\rm sin}\,\theta$ and ${\rm cos}\,\theta$ by their limiting values, 
$1/\beta$ and $\sqrt{\beta^2-1}/\beta$, respectively. We further note that
\hbox{${\cal D} \rightarrow 2\alpha\sqrt{\beta^2 - 1}$}. After differentiating
the asymptotic forms of $f_0^\prime$ and $g_{-2}$ with respect to $\theta$, we 
derive simplified, limiting forms of equations~(\ref{eqn:rSOu}) and 
(\ref{eqn:thetaSOu}). Expressed using $\alpha$ as the independent variable, 
these are
\begin{eqnarray}  
\label{eqn:rSOas}
&-&m\,\beta\,\zeta\,\alpha^{-m-1} \,\,+\,\, 
n\,\gamma\,\sqrt{\beta^2 - 1}\,\,\alpha^{-n-1}
\,\nonumber\\\, &=&-\frac{\beta^2}{4\,(\beta^2-1)}\,\,\alpha^{-1}
\\
\label{eqn:thetaSOas}
&&\beta\,\zeta\,\alpha^{-m} \,\,+\,\,
2\,(1-n)\,\gamma\,\sqrt{\beta^2-1}\,\,\alpha^{-n}
\,\nonumber\\\, &=&\frac{1}{2\,\sqrt{\beta^2-1}}\,\,\alpha^{-2} \,\,.
\end{eqnarray}
In deriving these equations, we have retained only the most rapidly diverging 
terms on the righthand sides of equations~(\ref{eqn:rSOu}) and 
(\ref{eqn:thetaSOu}). We have also dropped terms proportional to $\alpha^{-m}$ 
and  $\alpha^{-n}$ in the lefthand side of equation~(\ref{eqn:rSOu}), since 
both terms are dominated, in the small-$\alpha$ limit, by the two we have kept.

We next consider the $\beta$-dependence of $m$ and $n$. The righthand side of 
equation~(\ref{eqn:thetaSOas}) is proportional to $\alpha^{-2}$. For this 
equation to balance as $\alpha$ diminishes, one lefthand term could also 
diverge as $\alpha^{-2}$, and the other more slowly. However, if this were the 
case, i.e., if either \hbox{$m = 2 > n$} or \hbox{$n = 2 > m$}, then 
equation~(\ref{eqn:rSOas}) would be unbalanced. It is also posssible that the 
two lefthand terms in (\ref{eqn:thetaSOas}) diverge 
more
rapidly than 
$\alpha^{-2}$, balancing each other. We would then have \hbox{$m = n > 2$}. 
Finally, all three terms in equation~(\ref{eqn:thetaSOas}) could diverge at
the same rate: \hbox{$m = n = 2$}.

Assuming provisionally that \hbox{$m = n > 2$}, then we may ignore the 
righthand terms of both equations~(\ref{eqn:rSOas}) and (\ref{eqn:thetaSOas})
in the asymptotic limit. After dividing out all terms containing $\alpha$, the 
two equations reduce to
\begin{eqnarray*}
-\beta\,\zeta \,+\, \gamma\,\sqrt{\beta^2-1} \,&=&\, 0 \\
\beta\,\zeta \,+\, 2\,(1-n)\,\gamma\,\sqrt{\beta^2-1} \,&=&\, 0\,\,.
\end{eqnarray*}
Adding these two yields
\begin{equation*}
(3-2n)\,\gamma\,\sqrt{\beta^2-1} \,=\, 0 \,\,,
\end{equation*}
which implies that \hbox{$n\,=\,3/2$}. Since we assumed \hbox{$n > 2$} at the
outset, our original hypothesis, \hbox{$m = n > 2$}, was incorrect.

We have established that \hbox{$m = n = 2$}. Equations~(\ref{eqn:rSOas}) and 
(\ref{eqn:thetaSOas}) now reduce asymptotically to
\begin{eqnarray*}
-\beta\,\zeta \,+\, \gamma\,\sqrt{\beta^2-1} \,&=&\, 0 \\
\beta\,\zeta \,-\, 2\,\gamma\,\sqrt{\beta^2-1} \,&=&\, 
\frac{1}{2\,\sqrt{\beta^2-1}} \,\,,
\end{eqnarray*} 
which have the unique solution
\begin{eqnarray*}
\zeta \,&=&\, \frac{-1}{2\,\beta\,\sqrt{\beta^2-1}} \\ 
\gamma \,&=&\ \frac{-1}{2\,(\beta^2-1)} \,\,.
\end{eqnarray*}

Consider next the region just upstream from the Mach cone. Here, we may assume
the same asymptotic forms for $f_0^\prime$ and $g_{-2}$ as in 
equations~(\ref{eqn:fSOas}) and (\ref{eqn:gSOas}), provided the independent
variable $\alpha$ remains positive: 
\hbox{$\alpha\,\equiv\,\theta\,-\,\theta_M$}. Inserting these functional
forms into the second-order equations~(\ref{eqn:rSOi}) and (\ref{eqn:thetaSOi}),
we note that $d\alpha/d\theta$ changes sign from -1 to +1, and that
$\cal E$ now approaches \hbox{$2\,\alpha\,\sqrt{\beta^2-1}$} near the cone. We
then derive equations analogous to (\ref{eqn:rSOas}) and (\ref{eqn:thetaSOas}),
which may again be solved in the near-cone limit. After applying similar
reasoning to the two regions surrounding the anti-Mach cone, we arrive at the
following asymptotic forms for $f_0^\prime$ and $g_{-2}$ near $\theta_M$ and 
$\theta_M^\prime$:
\begin{align}\label{eqn:fas}
\begin{split}
f_0^\prime \,&=\, \frac{-1}{2\beta\sqrt{\beta^2-1}\,
\left(\theta_M -\theta\right)^2}  
\qquad{\theta\,\la\,\theta_M} \\
&=\, \frac{C^2}{2\beta\sqrt{\beta^2-1}\,
\left(\theta - \theta_M\right)^2} 
\qquad{\theta\,\ga\,\theta_M} \\
&=\, \frac{-C^2}{2\beta\sqrt{\beta^2-1}\,
\left(\theta_M^\prime - \theta\right)^2} 
\qquad{\theta\,\la\,\theta_M^\prime} \\
&=\, \frac{1}{2\beta\sqrt{\beta^2-1}\,
\left(\theta -\theta_M^\prime\right)^2} 
\qquad{\theta\,\ga\,\theta_M^\prime} \,\,.
\end{split}  
\end{align}
and
\begin{align}\label{eqn:gas}
\begin{split}
g_{-2} \,&=\, \frac{-1}{2\,\left(\beta^2-1\right)\,
\left(\theta_M -\theta\right)^2}  
\ \qquad{\theta\,\la\,\theta_M} \\
&=\, \frac{C^2}{2\,\left(\beta^2-1\right)\,
\left(\theta - \theta_M\right)^2} 
\ \qquad{\theta\,\ga\,\theta_M} \\
&=\, \frac{C^2}{2\,\left(\beta^2-1\right)\,
\left(\theta_M^\prime - \theta\right)^2} 
\ \qquad{\theta\,\la\,\theta_M^\prime} \\
&=\, \frac{-1}{2\,\left(\beta^2-1\right)\,
\left(\theta -\theta_M^\prime\right)^2} 
\ \qquad{\theta\,\ga\,\theta_M^\prime} \,\,.
\end{split}  
\end{align}

Obtaining $f_0$ requires that we integrate $f_0^\prime$ over $\theta$. Starting
at the upstream axis, with \hbox{$f_0 (\pi)\,=\,0$}, the integrated 
$f_0 (\theta)$ diverges to positive infinity as $\theta$ approaches 
$\theta_M^\prime$. However, the stream function has a direct physical meaning as
the mass transfer rate into a surface of revolution. Since this rate is finite 
for any $\theta$, the function $f_0^\prime (\theta)$ must be integrable across
the anti-Mach cone. That is, the 
upward
divergence of $f_0^\prime$ 
for \hbox{$\theta\,\ga\,\theta_M^\prime$} must be cancelled by a matching 
downward
divergence for \hbox{$\theta\,\la\,\theta_M^\prime$}. A similar 
antisymmetry of the divergences must occur at the Mach cone, 
\hbox{$\theta\,=\,\theta_M$} (see Fig.~\ref{fig:diverge}). Inspection of 
equation~(\ref{eqn:fas}) shows that this requirement forces $C^2$ to be unity. 
The parameter $C$ itself is either $+1$ or $-1$, independent of $\beta$. We 
will demonstrate presently that the positive solution is the physically 
relevant one. 

\subsection{Vorticity}

The second-order flow must be irrotational, as we argued in Section~3.3. 
Starting with equation~(\ref{eqn:Vorphi}), we again replace the velocity 
components by $\psi$ and $\rho$, and develop the latter in our perturbation 
series. Equating coefficients of $r^{-2}$ tests irrotationality in the 
second-order flow. Specifically, we require
\begin{eqnarray}
&\big(&\!-f_2^\prime\,g_{-1}^2 \,+\,f_2^\prime\,g_{-2} \,+
\,f_1^\prime\,g_{-1}\,-\,f_0^\prime \big)\ {\rm cot}\,\theta \,\,\nonumber\\
&+&f_2^{\prime\prime}\,g_{-1}^2 \,+\,2\,f_2^\prime\,g_{-1}\,g_{-1}^\prime \,-\,
f_2^{\prime\prime}\,g_{-2}\nonumber\\ 
\,&-&\,f_2^\prime\,g_{-2}^\prime \,-\,
f_1^{\prime\prime}\,g_{-1} \,-\, f_1^\prime\,g_{-1}^\prime \,+\, f_0^{\prime\prime}\,\nonumber
\\
&=&2\,f_2\,g_{-2}^2 \,-\,2\,f_2\,g_{-2} \,-\,f_1\,g_{-1} \,\,.
\label{eqn:irr}
\end{eqnarray}
Within the upstream and downstream regions, equations~(\ref{eqn:fFOup}) and
(\ref{eqn:gFOup}) give us $f_1$ and $g_{-1}$, respectively. Substituting these 
functions into equation~(\ref{eqn:irr}) then yields the condition of 
irrotationality: 
\begin{eqnarray}\label{irru}
\beta\,f_0^{\prime\prime} \,-\,\beta\,{\rm cot}\theta\,f_0^\prime \,-\,
\beta^2\,{\rm sin}\,\theta\,{\rm cos}\,\theta\,g_{-2}^\prime\, \nonumber\\+\,
2\,\beta^2\,{\rm sin}^2\,\theta\,g_{-2} \,=\,
\frac{1}{\cal D} \,-\, \frac{1}{\sqrt{\cal D}} \,\,.
\end{eqnarray}
Finally, we may combine this last relation with equation~(\ref{eqn:rSOu}), 
the $r$-component of Euler's equation, to obtain 
\begin{eqnarray}\label{eqn:irrusimple}
-2\,\beta\,{\rm cot}\,\theta\,f_0^\prime \,+\,
2\,\left(\beta^2\,-\,1\right) g_{-2}    
\,\nonumber\\=\,\frac{2}{\cal D} \,-\,
\frac{4}{\sqrt{\cal D}} \,+\,
\frac{2}{1\,+\,\sqrt{\cal D}} \,\,.
\end{eqnarray}
This is identical to equation~(I.64), and will prove useful when we evaluate the
force.

Turning to the intermediate region, we need to use equations~(\ref{eqn:fFOiC2})
and (\ref{eqn:gFOiC2}) for $f_1$, $g_{-1}$, and their derivatives. In both 
equations, the unknown parameter $C$ appears. Substitution into 
equation~(\ref{eqn:irr}) above yields the condition for irrotationality:
\begin{eqnarray}
\beta\,f_0^{\prime\prime} \,-\,\beta\,{\rm cot}\theta\,f_0^\prime \,-\,
\beta^2\,{\rm sin}\,\theta\,{\rm cos}\,\theta\,g_{-2}^\prime\,\nonumber\\+\,
2\,\beta^2\,{\rm sin}^2\,\theta\,g_{-2} \,
=\,
\frac{C^2}{\cal E} \,-\, \frac{C}{\sqrt{\cal E}} 
\,
\nonumber\\=\, \frac{1}{\cal E} \,-\, \frac{C}{\sqrt{\cal E}} \,\,.
\end{eqnarray}
In the last form of this equation, we have used the fact that 
\hbox{$C^2\,=\,1$}. Combination with the Euler equation~(\ref{eqn:rSOu}) gives 
\begin{eqnarray}\label{eqn:irrIsimple}
&-&2\,\beta\,{\rm cot}\,\theta\,f_0^\prime \,+\,
2\,\left(\beta^2\,-\,1\right) g_{-2}    
\,\nonumber \\ 
&=&  \,\frac{2\,C^2}{\cal E} \,-\, 
\frac{4\,C}{{\cal E}^{1/2}} \,+\,
\frac{1-C^2}{\beta^2\,{\rm sin}^2\,\theta}\,+\,
\frac{2\,C\,{\cal E}^{1/2}}{\beta^2\,{\rm sin}^2\,\theta} \nonumber
 \\ &=&\,\frac{2}{\cal E} \,-\,
\frac{4\,C}{{\cal E}^{1/2}} \,+\,
\frac{2\,C\,{\cal E}^{1/2}}{\beta^2\,{\rm sin}^2\,\theta} \,\,,
\end{eqnarray}
which will again aid in the force evaluation.

We emphasize a key difference between equations~(\ref{eqn:irrusimple}) and
(\ref{eqn:irrIsimple}). Because the upstream and downstream regions join 
smoothly onto the uniform background, the flow in both is guaranteed to be 
irrotational to all orders. In our numerical determination of the flow, to be 
described in Section~4.5 below, we verified that 
equation~(\ref{eqn:irrusimple}) indeed holds numerically both upstream and 
downstream. In contrast, we need to 
impose
equation~(\ref{eqn:irrIsimple}), or an equivalent condition, to obtain a 
physically acceptable solution in the intermediate region.

Enforcing irrotationality in the intermediate region requires only that we 
impose this condition along any interior cone, i.e., at a $\theta$-value such 
that \hbox{$\theta_M \,<\,\theta\,<\,\theta_M^\prime$}. To see why, return to
Euler's equation, in the form given by equation~(\ref{eqn:VorEuler}).
Projecting this vector equation along the $r$-direction gives
\begin{equation*}
u_\theta\ \omega_\phi \,=\, \frac{\partial B}{\partial r} \,\,.
\end{equation*} 
This relation holds at fixed $\theta$ and $\phi$. Since the flow is 
axisymmetric, the equation also holds at 
all
$\phi$-values, i.e., along 
a cone. If $\omega_\phi$ vanishes along such a cone, then $B$ is constant on 
that surface: \hbox{$\partial B/\partial r\,=\,0$}. But we already know that 
$B$ does not vary along any streamline. As Figure~\ref{fig:vort} illustrates 
schematically, the additional constraint tells us that the $B$-values 
characterizing each streamline are identical, i.e., $B$ is a true spatial 
constant. Thus, \hbox{$\bnabla B\,=\,0$} and, from 
equation~(\ref{eqn:VorEuler}), the flow is irrotational.

In practice, we apply \hbox{$\omega_\phi\,=\,0$} at \hbox{$\theta\,=\,\pi/2$}, 
the equatorial plane of the flow. The irrotationality condition, expressed as 
equation~(\ref{eqn:irrIsimple}), implies that there is a unique (but 
$\beta$-dependent) value of $g_{-2}$ at that angle:
\begin{eqnarray}\label{eqn:attractor}
g_{-2}\,(\pi/2)
 \,&=&\, \frac{2}{\big(\beta^2 - 1\big)^2} \,-\,
\frac{2\,C}{\big(\beta^2 - 1\big)^{3/2}} \,\nonumber\\&&+\,
\frac{C}{\beta\,\big(\beta^2 - 1\big)^{1/2}} \,\,.
\end{eqnarray}
When we find the flow numerically in the intermediate region, we will impose
equation~(\ref{eqn:attractor}) as an initial condition. The resulting flow is 
then irrotational. That is, all solutions of Euler's equations~(\ref{eqn:rSOi})
and (\ref{eqn:thetaSOi}) also obey  
equation~(\ref{eqn:irrIsimple}).

\subsection{Mass Accretion}

It is only in the second-order flow that mass accretion onto the central object
appears. Moreover, the function $f_0 (\theta)$, evaluated on the downstream
axis, gives the actual accretion rate. Higher-order approximations to the flow
contribute no additional information in this regard. We now review the 
argument, first advanced in Paper~I, and apply it to the present, supersonic, 
case.

Consider the net transfer rate of mass into a sphere of radius $r$. In a 
steady-state flow, this rate is the same through any closed surface; we simply
use a sphere for convenience. The dimensional result is
\begin{align}
\begin{split}
\label{eqn:mdotint}
{\dot M} \,&=\, -2\,\pi\!\int_0^\pi\!\!\rho\,u_r\,r^2\,{\rm sin}\,\theta\,
\,d\theta \\
&=\, 2\,\pi\,\psi (r,0) \,\,.
\end{split}
\end{align}
Here we have substituted equation~(\ref{eqn:urpsi}) for $u_r$ and utilized the 
normalization for the stream function, \hbox{$\psi (r,\pi)\,=\,0$}.\footnote
{By symmetry, the upstream axis coincides with a streamline, so that 
$\psi (r,\pi)$ is a constant. Such an additive constant does not effect any 
physical result, so we have set it to zero.} We have also implicitly assumed 
that $\rho$ and $u_r$ are smooth functions, despite the divergences arising in 
their perturbation expansions.

If we identify \hbox{$2\pi\rho_0 c_s r_s^2$} as our fiducial mass accretion 
rate, then the non-dimensional counterpart of the last equation becomes   
\begin{equation*}
{\dot M} \,=\, \psi (r,0) \,\,.
\end{equation*}
Substitution of the series expansion for $\psi (r,0)$ in 
equation~(\ref{eqn:exppsi}) and application of the boundary condition that 
\hbox{$f_i (0)\,=\,0$} for \hbox{$i\,=\,1,-1,-2$}, etc. leads to 
\begin{equation}\label{eqn:mdotf} 
{\dot M} \,=\, f_0 (0) \,\,.
\end{equation}
It is noteworthy that $\dot{M}$ depends only on a coefficient from the second-order
    expansion; this fact calls for a more physical explanation. One reason is that
    the streamlines of the first-order flow are symmetric, so that accretion does 
    not occur to this approximation. Secondly, the mass flux $\rho\, u_r$, when expanded
    using approximations higher than second-order, generates terms that fall off
    faster than $r^{-2}$. After integrating these terms over a large bounding
    sphere and taking the large-$r$ limit, they do not contribute to $\dot{M}$. In any
    event, equation (49) underscores the fact that the function 
\hbox{$f_0^\prime (\theta)$} must be integrable across both Mach cones, despite
its divergent behavior at the cones themselves.

We may now use our second-order equations to obtain a useful relation between
$f_0 (0)$, and hence the mass accretion rate, and the flow density. Consider 
first the upstream and downstream regions. The lefthand side of 
equation~(\ref{eqn:thetaSOu}) is a perfect derivative:
\begin{eqnarray*}
-\beta\,f_0^\prime \,+\,{\cal D}\,g_{-2}^\prime \,-\,
2\,\beta^2\,{\rm sin}\,\theta\,{\rm cos}\,\theta\,g_{-2} \,\\=\,
{{d\phantom r}\over{d\theta}}
\left[-\beta\,f_0 \,+\,\left(1\,-\,\beta^2\,{\rm sin}\,\theta\right)\,g_{-2}
\right]\,\,.
\end{eqnarray*}
In the righthand side of the same equation, we note that $\cal D$ is an
even
function, in the sense that 
\hbox{${\cal D} (\theta) \,=\, {\cal D} (\pi -\theta)$}. Since 
${\rm cos}\,\theta$ is an 
odd
function, 
\hbox{${\rm cos}\,\theta \,=\, -{\rm cos}\,(\pi - \theta)$},
the entire righthand side of equation~(\ref{eqn:thetaSOu}) is odd.

Within the intermediate region, the lefthand side of 
equation~(\ref{eqn:thetaSOi}) is
\begin{eqnarray*}
-\beta\,f_0^\prime\,-\,{\cal E}\,g_{-2}^\prime \,-\,
2\,\beta^2\,{\rm sin}\,\theta\,{\rm cos}\,\theta\,g_{-2} \,\\=\,
{{d\phantom r}\over{d\theta}}
\left[-\beta\,f_0 \,+\,\left(1\,-\,\beta^2\,{\rm sin}^2\,\theta\right)\,g_{-2}
\right] \,\,.
\end{eqnarray*}
Again, the righthand side of equation~(\ref{eqn:thetaSOi}) is odd. Now the 
integral of an odd function from the downstream to the upstream axes vanishes. 
Thus, if we integrate equation~(\ref{eqn:thetaSOu}) from \hbox{$\theta = 0$} 
to $\theta_M$, equation~(\ref{eqn:thetaSOi}) from $\theta_M$ to 
$\theta_M^\prime$, and equation~(\ref{eqn:thetaSOu}) from $\theta_M^\prime$ to 
$\pi$, we find that
\begin{equation*}
-\beta\,f_0 (0) \,+\, g_{-2} (0) \,=\,
-\beta\,f_0 (\pi) \,+\, g_{-2} (\pi) \,\,.
\end{equation*}
Using \hbox{$f_0 (\pi) \,=\,0$}, we have the desired result:
\begin{equation}\label{eqn:fg}
f_0 (0) \,=\,\frac{g_{-2} (0) \,-\, g_{-2} (\pi)}{\beta} \,\,.
\end{equation}
The same relation appeared in the subsonic study as equation~(I.46).

\subsection{Numerical Solution}

\subsubsection{Evaluation of C}

Our description of the flow is clearly incomplete until we identify the still 
unknown parameter $C$, and not just its absolute value. We again employ 
physical reasoning. The value of $C$ affects the shape of the streamlines, even
in the first-order flow. Only one shape is reasonable dynamically. 
Figure~\ref{fig:fos11} shows streamlines of the first-order flow for the case 
\hbox{$\beta\,=\,1.1$}. That is, we plot contours of constant
\hbox{$f_2\,r^2\,+\,f_1\,r$}. In the upstream and donwstream regions,
$f_1$ is taken from equation~(\ref{eqn:fFOup}). For the intermediate region,
we use equation~(\ref{eqn:fFOiC2}), with both \hbox{$C\,=\,+1$} (solid
curves) and \hbox{$C\,=\,-1$} (dashed curves). For either choice of the 
parameter, we see that all first-order streamlines are symmetric about the
equatorial plane, as we found in the subsonic flow (see, e. g., Fig.~I.2). The kinks at both Mach cones result from the crudeness of this first-order
    approximation, and would disappear in successively more accurate treatments.

The solid curves in Figure~\ref{fig:fos11} bow inward, toward the central 
mass, both upstream of the anti-Mach cone and inside it. Such inward turning 
results naturally from the gravitational pull of the mass. Eventually, the 
concurrent rise in density and pressure pushes the gas outward. If we were 
instead to choose \hbox{$C\,=\,-1$}, the streamlines in the intermediate 
region would bow outward, which is not expected. 

We may also consider the matter more quantitatively by examining the velocity
component $u_R$. Here $R$ is the cylindrical radius: 
\hbox{$R\,\equiv\,r\,{\rm sin}\,\theta$}. In the far-field limit, $u_R$ tends
to zero, since the background flow is in the $z$-direction. Closer to the
equatorial plane, but still upstream, $u_R$ should be slightly negative.
Downstream, this velocity component should reverse sign as each streamline
rejoins the background flow. Starting with the expreessions for $f_1$ and 
$g_{-1}$ in the intermediate region, we first calculate $u_r$ and $u_\theta$ from 
equations~(\ref{eqn:urpsi}) and (\ref{eqn:uthetapsi}), and then obtain 
\hbox{$u_R\,\equiv\,u_r\,{\rm sin}\,\theta+\,u_\theta\,{\rm cos}\,\theta$}. We
find that 
\begin{equation*}
u_R\,=\,
\left[\frac{C\,-\,{\cal E}^{1/2}}{r\,\beta\,{\cal E}^{1/2}}\right]
{\rm cot}\,\theta \,\,.
\end{equation*}
Upstream from the mass \hbox{$\left(\theta\,>\,\pi/2\right)$}, we have
\hbox{${\rm cot}\,\theta\,>\,0$}. Thus, if \hbox{$C\,=\,-1$}, then
\hbox{$u_R\,>\,0$} in this direction. Conversely, \hbox{$u_R\,<\,0$} 
downstream from the mass. This behavior is contrary to the physically 
reasonable one, so we conclude that \hbox{$C\,=\,+1$}.

\subsubsection{Integration Scheme}

Determining the second-order flow requires that we numerically integrate
equations~(\ref{eqn:rSOu}) and (\ref{eqn:thetaSOu}) in the upstream and 
downstream regions, and equations~(\ref{eqn:rSOi}) and (\ref{eqn:thetaSOi}),
with \hbox{$C\,=\,+1$}, in the intermediate region. In all cases, we need to 
specify appropriate boundary values of $f_0^\prime$ and $g_{-2}$. We then perform
an additional integration to obtain $f_0$ and thereby the streamlines.

One of our boundary conditions is the value of $f_0 (0)$. As we have seen, this
quantity is also the mass accretion rate $\dot M$. There is still no 
fundamental theory to supply this rate, except at \hbox{$\beta\,=\,0$}
\citep{b52}, and in the hypersonic \hbox{$(\beta\,\gg\,1)$} limit
\citep{hl39, bh44}. Extending the work of \citet{b52}, \citet{mt09} suggested 
an interpolation formula that both respects the analytic limits and agrees 
reasonably well with numerical simulations. As we did in Paper~I, we adopt this
prescription, which is
\begin{align}\label{eqn:throop}
\begin{split}
{\dot M} (\beta)\,&=\,\frac
{2 \left(\lambda^2\,+\,\beta^2\right)^{1/2}}
{\left(1\,+\,\beta^2\right)^2} \\
&=\,f_0 (0) \,\,.
\end{split}
\end{align}
Here, \hbox{$2\,\lambda\,=\,{\rm e}^{3/2}/2\,=\,2.24$} is the analytic value of
${\dot M} (0)$ in the isothermal case \citep{b52}.
 
We are now in a position to outline our numerical procedure. Starting at the
upstream axis, \hbox{$\theta\,=\,\pi$}, we set 
\hbox{$f_0 (\pi)\,=\,f_0^\prime (\pi)\,=\,0$}. We provisionally treat
$g_{-2} (\pi)$ as a free parameter, to be determined later in order to ensure
smoothness of the flow across the Mach cones, as we describe in more detail
below. For any selected $g_{-2} (\pi)$, we may then integrate 
equations~(\ref{eqn:rSOu}) and (\ref{eqn:thetaSOu}) to the anti-Mach cone, 
\hbox{$\theta\,=\,\theta_M^\prime$}. Simultaneous integration of $f_0^\prime$ 
yields $f_0$ itself.

But knowledge of both $g_{-2} (\pi)$ and $f_0 (0)$ also gives us $g_{-2} (0)$,
according to equation~(\ref{eqn:fg}). Since \hbox{$f_0^\prime (0)\,=\,0$}, we 
may again integrate the second-order equations~equations~(\ref{eqn:rSOu}) and 
(\ref{eqn:thetaSOu}), along with $f_0^\prime$, from \hbox{$\theta\,=\,0$} to 
\hbox{$\theta\,=\,\theta_M$}. At this point, we have established a 
one-parameter family of flows covering both the upstream and downstream 
regions. For any value of $g_{-2} (\pi)$, both flows are irrotational, as 
required physically.

Turning to the intermediate region, we begin at the equatorial plane,
\hbox{$\theta\,=\,\pi/2$}. We use equation~(\ref{eqn:attractor}) for
\hbox{$g_{-2} (\pi/2)$}, again setting \hbox{$C\,=\,+1$}. Our free parameter is
now $f_0^\prime (\pi/2)$. For any value of this quantity, we integrate 
equations~(\ref{eqn:rSOi}) and (\ref{eqn:thetaSOi}) away from the plane in 
both directions until we come to the Mach cones. We have then established the 
run of $f_0^\prime$ and $g_{-2}$ throughout the intermediate region. Again, the 
flow is irrotational for any value of 
$f_0^\prime (\pi/2)$.\footnote{On a related issue, not only do the flows in 
all regions have spatially uniform values of the Bernoulli function $B$, but 
these values match: \hbox{$B\,=\,\beta^2/2$}. To see this, we may expand 
equation~(\ref{eqn:Bf}) for $B$, using our perturbation series. The 
lowest-order ($r$-independent) terms sum to $\beta^2/2$ in all regions, while 
the higher-order terms sum to zero.}

\subsubsection{Enforcing Smoothness}

Within the intermediate region, we only know $f_0$ up to a constant of
integration, which has yet to be fixed. In addition, we have not yet determined
our two free parameters, $f_0^\prime (\pi/2)$ and $g_{-2} (\pi)$. All three
quantities are specified by requiring that the flow be smoothly varying. If 
$f_0^\prime$ and $g_{-2}$ were well-defined everywhere, this task would be
straightforward. For example, we could tune one parameter until $g_{-2}$,
as calculated in the downstream region, matched the intermediate $g_{-2}$ 
at the Mach cone. However, both $f_0^\prime$ and $g_{-2}$ diverge at the cones. 
Hence, we can only require smoothness outside some finite region surrounding 
each cone.

We therefore stopped each integration at a point where the divergent behavior
begins to dominate. For example, when integrating from 
\hbox{$\theta\,=\,\pi$}, we examined the ratio
\hbox{$\alpha \,\equiv\,\vert{(\pi - \theta)\,
f_0^{\prime\prime}}\,/\,{f_0^\prime}\vert$}. 
Since \hbox{$f_0^\prime (\pi)\,=\,0$}, this quantity is close to unity for 
\hbox{$\theta\,\lesssim\,\pi$}. However, $\alpha$ climbs sharply near 
$\theta_M^\prime$. In practice, we stopped the integration at $\alpha\,=\,10$. We
adopted equivalent criteria in the other regions. thus establishing boundaries 
for the well-behaved flow. Finally, we chose $f_0^\prime (\pi/2)$ and 
$g_{-2} (\pi)$ so that $f_0$ and $g_{-2}$ matched as closely as possible across 
these boundaries.

A convenient measure of the goodness of fit is
\begin{equation}\label{eqn:chi}
\chi^2 \,\equiv\, \Delta g_{-2}^2 \left(\theta_M^\prime\right) \,+\,
\Delta f_0^2 \left(\theta_M\right) \,+\,
\Delta g_{-2}^2 \left(\theta_M\right) \,\,.
\end{equation} 
Here, \hbox{$\Delta g_{-2} \left(\theta_M^\prime\right)$} is difference of
$g_{-2}$ across the anti-Mach cone, and the two other quantities are analogously
defined. We separately enforced
\hbox{$\Delta f_0 \left(\theta_M^\prime\right)\,=\,0$} by adding the appropriate 
constant to the numerical integration of $f_0^\prime$ in the intermediate region.
As we varied both $f_0^\prime (\pi/2)$ and $g_{-2} (\pi)$, the quantity $\chi$
reached a well-defined minimum. Figure~\ref{fig:contour} shows $\chi$-contours 
for the case \hbox{$\beta \,=\,1.1$}. The optimal values of 
$f_0^\prime (\pi/2)$ and $g_{-2} (\pi)$, shown here by the dot within in circle,
were insensitive to our choice of $\alpha$.

\subsubsection{Sample Results}

With the numerical procedure in hand, we could determine the second-order flow
for any desired $\beta$. Figure~\ref{fig:sos11} displays streamlines for 
$\beta\,=\,1.1$. Even in this modestly supersonic case, the two Mach cones 
depart substantially from the equatorial plane \hbox{$(\theta_M\,=\,65^\circ)$},
and the isodensity contours are a set of nearly vertical lines that bow inward
slightly toward the mass. The shaded interiors of the wedges straddling each 
Mach cone represent the excluded sectors within which the divergent behavior 
dominates. We previously indicated the excluded region surrounding the Mach 
cone by the shading in Figure~\ref{fig:diverge}.

Returning to Figure~\ref{fig:sos11}, we notice that the streamlines in the 
intermediate region have the expected concavity, but are not precisely 
symmetric across the equatorial plane. 
For example, a cone with a half angle of $\theta = 50^\circ$ cuts the outermost streamlines shown at a radius that is 1\% closer downstream than upstream.  
Notice also how the innermost 
streamlines join onto the central mass. The surface of revolution they generate
encloses the full $\dot M$, and the figure illustrates that mass accretion 
occurs in this order of approximation. However, we cannot obtain the detailed 
behavior of the flow as it joins onto the mass, since our perturbation series 
is only valid well outside \hbox{$r\,=\,1$}, the boundary indicated here as 
the central, dashed circle.

Figure~\ref{fig:gfp11} shows the angular variation of $f_0^\prime$ (dashed
curve) and $g_{-2}$ (dotted curve) for this same $\beta$-value. As in 
Figure~\ref{fig:sos11}, the shaded regions mark those sectors where both 
variables diverge. By design, the coefficients diverge antisymmetrically as
either cone is approached. This behavior is not evident in the figure, which
appears to show a symmetric divergence. In more detail, both $f_0^\prime$ and
$g_{-2}$ first rise when approaching the Mach cone from downstream, then reach a
peak and plunge downward (recall Fig.~\ref{fig:diverge}). They exhibit 
analogous behavior upstream of the anti-Mach cone.

The solid curve in Figure~\ref{fig:gfp11} shows the integrated $f_0$. The
upstream and downstream values of this coefficient match exactly at the 
anti-Mach cone. We forced this match by adjusting the integration constant
within the intermediate region. The values of $f_0$ on either side of the Mach
cone have a ratio of 1.6. This ratio depends both on our choices of
$f^\prime (\pi/2)$ and $g_{-2} (\pi/2)$, and on the precise manner by
which we excise the divergent region surrounding the Mach cone. Within our
scheme, the mismatch is a weak function of the parameter $\alpha$.  

As an additional example, Figure~\ref{fig:sos20} displays streamlines of the
second-order flow for \hbox{$\beta\,=\,2.0$}. Here, both cones are even
more removed from the equatorial plane (\hbox{$\theta_M \,=\, 30^\circ$}), and 
the fluid elements are nearly following straight-line trajectories. 
The innermost pair shown here nevertheless still bends to reach the origin.
Quantitatively, the fluid is affected significantly by gravity if it passes
within the accretion radius $r_{\rm acc}$, which varies as $\beta^{-2}$. Notice 
also that the figure of revolution generated by the innermost streamlines is 
much narrower than in Figure~\ref{fig:sos11}. This narrowing reflects the
fact that $\dot M$ decreases steeply with $\beta$ as the Mach number climbs
significantly above unity.

We remind the reader that both Figure~\ref{fig:sos11} and \ref{fig:sos20} are just approximations to the actual flow. 
The calculated flow would become
more accurate at any fixed $ r \gg r_{\rm s}$ with the inclusion of higher-order
terms in the perturbation series. However, the mathematical
divergences at the Mach and anti-Mach cones would remain. Because of
these divergences, streamlines tend to bend unphysically toward both
cones. Our procedure above was designed to prevent these divergences
from unduly influencing the shapes of the
streamlines, at least to this order in the perturbation expansion.

\section{Friction Force}

\subsection{Integration of the Momentum Flux}

Our reconstruction of the streamlines has a degree of arbitrariness, 
necessitated by the near-cone divergence in our perturbation series. We will
now show that the force evaluation does not have this limitation, but is exact.
Dimensionally, the total rate at which $z$-momentum is transported into a 
sphere surrounding the mass is
\begin{eqnarray}\label{eqn:Fdim}
{\dot P} \,&=&\, -2\,\pi\!\int_0^\pi\!
              \rho\,u_r\,u_z\,r^2\,{\rm sin}\,\theta\ d\theta \,\,\nonumber\\&&-\,\,
2\,\pi\!\int_0^\pi\!
                 \rho\,c_s^2\,r^2\,{\rm cos}\,\theta\ {\rm sin}\,\theta\,\,
                 d\theta \,\,.
\end{eqnarray}
The first righthand term is the net advection of momentum across the surface,
while the second represents the static pressure acting on the sphere. If we
let our unit of force be \hbox{$2\pi\rho_0\,c_s^2\,r_s^2$}, then the
corresponding non-dimensional expression is
\begin{eqnarray}\label{eqn:Fnondim}
{\dot P}\,&=&\, -\int_0^\pi\!\rho\,u_r\,u_z\,r^2\,{\rm sin}\,\theta\ d\theta \,\nonumber\\ &&-\,
        \int_0^\pi\!\rho\,r^2\,{\rm cos}\,\theta\ {\rm sin}\,\theta\,\,
        d\theta \,\,.
\end{eqnarray}

The next step is to write all velocity components in terms of $\psi$ and $\rho$,
and then to expand the latter variables in their respective perturbation 
series. These operations produce terms proportional to $r^2$, $r^1$, $r^0$, 
etc. We are thus motivated to write
\begin{equation}\label{eqn:Fseries}
{\dot P} \,=\,{\dot P}_2\,r^2 \,+\,{\dot P}_1\,r^1\,+\,{\dot P}_0\,r^0
\,+\,...  \,\,\,.
\end{equation}
Since $\dot P$ is also the force on the mass, it should not depend on distance.
It is important, therefore, to check that both ${\dot P}_2$ and ${\dot P}_1$ 
vanish identically. This is indeed the case. We refer the reader to Appendix~A 
for details of the calculation. The terms in $\dot P$ that are proportional to 
negative powers of $r$ diminish at large distance and need not be calculated 
explicitly. For the remainder of this section, we focus on evaluating the term 
${\dot P}_0$.

The full expression for ${\dot P}_0$ may be written as
\begin{equation}
{\dot P}_0 \,=\, -\int_0^\pi \! \left({\cal F}_1\, +\, {\cal F}_2\right)\, 
d\theta \,\,,
\end{equation}  
where
\begin{eqnarray*}
{\cal F}_1 \,&\equiv&\, \left(1 + {\rm cos}^2\,\theta\right) \beta\,f_0^\prime
\,\,\nonumber\\ &&+\,\,\left(1 - \beta^2\right) {\rm sin}\,\theta\,{\rm cos}\,\theta\,\,g_{-2} 
\,\,\\&&+\,\, f_1\,f_1^\prime \,\,+\,\,\left(f_1^\prime\right)^2\,{\rm cot}\,\theta
\,\,,
\end{eqnarray*}
and  
\begin{eqnarray*}
{\cal F}_2 \,&\equiv&\, \beta^2\,{\rm sin}\,\theta\,{\rm cos}\,\theta\,\,g_{-1}^2
\,\,\\&&-\,\,\beta\,{\rm sin}\,\theta\,{\rm cos}\,\theta\,f_1\,g_{-1} \,\,\\&&-\,\,
\left(1+{\rm cos}^2\,\theta\right) \beta\,f_1^\prime\,g_{-1} \,\,.
\end{eqnarray*}

If we use equations~(\ref{eqn:fFOup}) and (\ref{eqn:gFOup}) for $f_1$ and
$g_{-1}$ in the downstream region, then we find that this contribution to
the full ${\dot P}_0$, which we label ${\dot P}_0^d$, is
\begin{eqnarray}\label{eqn:pdotd}
{\dot P}_0^d \,=\, -\int_0^{\theta_M} \!\!d\theta\left[\left(1-\beta^2\right) 
{\rm sin}\,\theta\,{\rm cos}\,\theta\,g_{-2} \,\right. \nonumber\\\left. +\, \beta  
\left(1 + {\rm cos}^2\,\theta\right)f_0^\prime\right] \,\,.
\end{eqnarray}
At this point, we may utilize the condition of irrotationality, expressed as
equation~(\ref{eqn:irrusimple}). Multiplying this latter equation by
\hbox{$({\rm sin}\,\theta\,{\rm cos}\,\theta /2)$} gives
\begin{eqnarray*}
-\left(1-\beta^2\right) {\rm sin}\,\theta\,{\rm cos}\,\theta\,\,g_{-2} \,-\,
\beta\,{\rm cos}^2\,\theta\,f_0^\prime \,\,\\=\,\, {\rm sin}\,
\theta\,{\rm cos}\,\theta
\left(\frac{1}{\cal D} \,-\, \frac{2}{\sqrt{\cal D}} \,+\,
\frac{1}{1+\sqrt{\cal D}}\right) \,\,.
\end{eqnarray*}

Substitution of this last relation into equation~(\ref{eqn:pdotd}) transforms 
the latter into
\begin{eqnarray*}
{\dot P}_0^d \,=\, 
\int_0^{\theta_M}\!\! d\theta\,\,{\rm sin}\,\theta\,{\rm cos}\,\theta
\left(\frac{1}{1+\sqrt{\cal D}}-\frac{2}{\sqrt{\cal D}}\right)
\,\\+\, \int_0^{\theta_M}\!\!d\theta\,\,\,
\frac{{\rm sin}\,\theta\,{\rm cos}\,\theta}{\cal D} \,-\,
\beta\!\int_0^{\theta_M}\!\!f_0^\prime\,d\theta \,\,.
\end{eqnarray*}
Of the three integrals on the righthand side, the first may be evaluated
analytically, while the second and third are divergent. After doing the
first integral, we have
\begin{eqnarray}\label{eqn:pdotdf}
{\dot P}_0^d \,&=&\,
\frac{-1 -{\rm ln}\,2}{\beta^2}
\,+\, \int_0^{\theta_M}\!\!d\theta\,\,\,
\frac{{\rm sin}\,\theta\,{\rm cos}\,\theta}{\cal D} \, \nonumber\\ &&-\,
\beta\!\int_0^{\theta_M}\!\!f_0^\prime\,d\theta \,\,.
\end{eqnarray}
Closely analogous reasoning for the upstream contribution, to be denoted
${\dot P}_0^u$, gives us 
\begin{eqnarray}\label{eqn:pdotuf}
{\dot P}_0^u \,&=&\,
\frac{1 +{\rm ln}\,2}{\beta^2}
\,+\, \int_{\theta_M^\prime}^\pi\!\!d\theta\,\,\,
\frac{{\rm sin}\,\theta\,{\rm cos}\,\theta}{\cal D} \,\nonumber\\&&-\,
\beta\!\int_{\theta_M^\prime}^\pi\!\!f_0^\prime\,d\theta \,\,.
\end{eqnarray}

For the intermediate region, we must now use 
equations~(\ref{eqn:fFOiC2}) and (\ref{eqn:gFOiC2}) for $f_1$ and $g_{-1}$, 
with \hbox{$C\,=\,+1$}. We find, for the contribution ${\dot P}_0^i$, an
equation analogous to (\ref{eqn:pdotd}):
\begin{eqnarray}\label{eqn:pdoti}
{\dot P}_0^i \,=\, -\int_{\theta_M}^{\theta_M^\prime} 
\!\!d\theta\left[\left(1-\beta^2\right) 
{\rm sin}\,\theta\,{\rm cos}\,\theta\,g_{-2} \,\right.\nonumber\\\left.+\, \beta  
\left(1 + {\rm cos}^2\,\theta\right)f_0^\prime\right] \,\,.
\end{eqnarray}
We again avail ourselves of irrotationality, now in the form of
equation~(\ref{eqn:irrIsimple}). Multiplication of this last equation by
\hbox{$({\rm sin}\,\theta\,{\rm cos}\,\theta /2)$} gives 
\begin{eqnarray*}
&-&\left(1-\beta^2\right) {\rm sin}\,\theta\,{\rm cos}\,\theta\,\,g_{-2} \,-\,
\beta\,{\rm cos}^2\,\theta\,f_0^\prime \,\,\\&=&\,\, {\rm sin}\,
\theta\,{\rm cos}\,\theta
\left(\frac{1}{\cal E} \,-\, \frac{2}{\sqrt{\cal E}} \,+\,
\frac{{\cal E}^{1/2}}{\beta^2\,{\rm sin}^2\,\theta}
\right) \,\,.
\end{eqnarray*}
Thus, equation~(\ref{eqn:pdoti}) becomes
\begin{eqnarray}\label{eqn:pdotit}
{\dot P}_0^i \,&=&\,\,
-\beta\!\int_{\theta_M}^{\theta_M^\prime}\!\!f_0^\prime\,d\theta \nonumber\\ &+&
\int_{\theta_M}^{\theta_M^\prime}\!\! d\theta\,\,{\rm sin}\,\theta\,{\rm cos}\,\theta
\left(
\frac{1}{\cal E} \, -\,
\frac{2}{{\cal E}^{1/2}} \, \right. \nonumber\\ && \left. +\,
\frac{{\cal E}^{1/2}}{\beta^2\,{\rm sin}^2\,\theta}
\right)
\end{eqnarray}

The integrand within the second righthand term of equation~(\ref{eqn:pdotit}) is
an odd function. Since the range of integration is symmetric about
\hbox{$\theta\,=\,\pi/2$}, this term vanishes, and we are left with
\begin{equation}\label{eqn:pdotif}
{\dot P}_0^i \,=\, 
-\beta\!\int_{\theta_M}^{\theta_M^\prime}\!\!f_0^\prime\,\,d\theta \,\,.
\end{equation}
Combining equations~(\ref{eqn:pdotdf}), (\ref{eqn:pdotuf}), and 
(\ref{eqn:pdotif}), we find
\begin{eqnarray}\label{eqn:pdotf}
{\dot P}_0 \,&=&\, {\dot P}_0^d \,+\,{\dot P}_0^u \,+\,{\dot P}_0^i \nonumber\\
&=&\,
\int_0^{\theta_M} \!\! d\theta\,\,\, 
\frac{{\rm sin}\,\theta\,{\rm cos}\,\theta}{\cal D}\,\,\nonumber\\&&+\,\,
\int_{\theta_M^\prime}^\pi \!\! d\theta\,\,\, 
\frac{{\rm sin}\,\theta\,{\rm cos}\,\theta}{\cal D}\,\,\nonumber\\&&-\,\
\beta\!\int_0^\pi \!\!f_0^\prime\,\,d\theta \,\,.
\end{eqnarray}
Within the second righthand integral, let \hbox{$\nu\,\equiv\,\pi\,-\,\theta$}.
Noting that \hbox{$\pi\,-\,\theta_M^\prime\,=\,\theta_M$}, we find
\begin{eqnarray*}
&&\int_{\theta_M^\prime}^\pi\!\!d\theta\,\,\,
\frac{{\rm sin}\,\theta\,{\rm cos}\,\theta}{1 - \beta^2\,{\rm sin}^2\,\theta}
\,=\\ &-&
\int_0^{\theta_M}\!\!d\nu\,\,\,
\frac{{\rm sin}\,\nu\,{\rm cos}\,\nu}{1 - \beta^2\,{\rm sin}^2\,\nu}
\,\,.
\end{eqnarray*}
Thus, the first and second righthand integrals of equation~(\ref{eqn:pdotf})
cancel, and we have
\begin{equation}
{\dot P}_0 \,=\, -\beta \int_0^\pi \!\! f_0^\prime \,\,d\theta \,=\,
\beta\,f_0 (0) \,\,.
\end{equation}
Using equation~(\ref{eqn:mdotf}), and identifying ${\dot P}_0$ as the friction
force $F$, we arrive at our central result:
\begin{equation}\label{eqn:Ff}
F \,=\, {\dot M}\, \beta \,\,.
\end{equation}

Equation~(\ref{eqn:Ff}) is identical to that derived in the subsonic case,
equation~(I.58). What we have demonstrated is that a single expression for the
force holds at all speeds. Given our assumption of a steady-state, laminar 
flow in the far field, this result is exact. Although the force involves the
factor $\dot M$, our expression holds even for a porous, non-accreting object,
such as a globular cluster traversing a galactic halo. The result also applies
to a mass that is expelling its own gas, e.g., a wind-emitting star. In that
case, the wind can only penetrate a limited distance before it shocks against 
the background flow. The dynamical friction force then acts on both the mass
and its trapped wind. Our only requirement is that the background gas well 
outside of $r_{\rm s}$ be undisturbed. 

This last stipulation is overly conservative. Mathematical convenience
originally motivated our choice of $r_{\rm s}$ as the characteristic length
scale.  For supersonic flows, however, streamlines deviate from the
background not inside $r_{\rm s}$, but instead inside $r_{\rm acc} \sim r_{\rm s}/\beta^2 < r_{\rm s}$. Numerical simulations show this effect \citep[e.g.,][]{s85,r96,l13}.  In conclusion, it is the flow outside
of $r_{\rm acc}$ that physically determines the force, regardless of the
domain of validity for our perturbation series.

For computational purposes, we may use in equation~(\ref{eqn:Ff}) the 
interpolation formula for $\dot M$ given in equation~(\ref{eqn:throop}):
\begin{equation}\label{eqn:Fthroop}
F \,=\, \frac
{2\,\beta \left(\lambda^2 \,+\, \beta^2\right)^{1/2}}
{\left(1\,+\,\beta^2\right)^2}  \,\,.
\end{equation}
We plot this relation in Figure~\ref{fig:force}. We also show, for 
comparison, the friction force derived by \citet{o99}, as given in her equations (14) and (15). For \hbox{$\beta\,>\,1$},
her analytic formula includes the factor 
$\alpha_\circ\,\equiv\,{\rm ln}\,\left(V t/R\right)$, where $t$ is the
time since the gravitational force is switched on. In the figure, we have
set \hbox{$\alpha_\circ\,=\,2$}. Notice that Ostriker's force diverges at 
\hbox{$\beta \,=\,1$}, while ours is continuous at all Mach numbers. 

\subsection{Comparison with Simulations} 

The numerical simulation of gas streaming past a stationary, gravitating mass 
has a long history. In an early work, \citet{h71} investigated a fluid with an
adiabatic index of \hbox{$\gamma\,=\,5/3$}, and with incident Mach numbers 
ranging from 0.6 to 2.4. A study more directly relevant to ours is that of 
\citet{s85}, whose suite of axisymmetric simulations included a 
\hbox{$\gamma\,=\,1.1$} gas with Mach numbers again up to 2.4. The authors 
calculated both contributions to the friction force $F$ -- the gravitational 
tug from the wake, determined by integration over the density around the mass, 
and the advection of linear momentum through the surface of this body, a force 
variously called the aerodynamic or hydrodynamic drag. Despite the nomenclature,
they found that this second contribution is actually a forward thrust, directed
oppositely to the gravitational force. Recall that external gas does not impact 
the mass directly, but misses it and enters from downstream, thus imparting 
momentum in the upstream direction.

The total force found by \citet{s85} is greater than ours. These authors
define a drag coefficient $c_d$ through the relation
\hbox{$F\,=\,2\,\pi\,c_d\,\rho_0\,G^2\,M^2/V^2$} and an effective accretion
radius $r_{\rm eff}$ through 
\hbox{${\dot M} \,=\, \pi\,r_{\rm eff}^2\,\rho_0\,V$}. It follows that
\begin{equation*}
\frac{F}{{\dot M}\,V} \,=\, \frac{c_d}{2} 
\left(\frac{r_{\rm acc}}{r_{\rm eff}}\right)^2  \,\,.
\end{equation*}
Using their Table~1 and Figure~A.9, we infer that, for \hbox{$\beta\,=\,0.6$},
1.4, and 2.4, \hbox{$F/{\dot M}\,V\,=\,1.0$}, 2.4, and 3.6, respectively. Note
that the radius $R$ of their accreting object varied with Mach number. It was
set to 0.1 times $r_{\rm acc}$, so that \hbox{$R\,=\,0.2\,r_s/\beta^2$}.

In a later study, \citet{sb99} simulated the flow of an isothermal gas around 
a mass whose physical size $R$ excceds $r_{\rm acc}$ by a factor of 20 or more. 
Under these conditions, the density enhancement in the wake is relatively mild. 
\citet{sb99} switched on the gravity from the central mass suddenly, as in the 
calculation of \citet{o99}. They found that their computed gravitational 
portion of the force closely matched Ostriker's analytical prediction. More 
recently, \citet{kk09}, who adopted \hbox{$\gamma\,=\,5/3$}, showed that this 
contribution to the force declines when $R$ falls well below $r_{\rm acc}$ (see 
their Fig.~14).

The most thorough numerical investigation of the flow pattern, mass accretion 
rate, and friction force for the isothermal case has been that of \citet{r96}. 
In this three-dimensional simulation of a \hbox{$\gamma\,=\,1.01$} gas, both 
the accretor size and Mach number were varied, the latter up to
\hbox{$\beta\,=\,10$}. A general result, which corroborated and extended those
of previous studies, was that the downstream region exhibited continuing
instability for relatively small masses \hbox{$\left(R\,\ll\,r_s\right)$} 
embedded in supersonic flow. These, of course, are just the conditions of most 
interest for our purposes. \citet{r96} reported time-averaged results for both 
$\dot M$ and $F$ in such cases. 

Table~1 displays the essential results found by \citet{r96}, both for his 
subsonic simulations \hbox{$(\beta\,=\,0.6)$} and supersonic ones. We have
only taken data from the runs in which the central mass was the smallest size.
Here, the radius was 0.02 times $r_{\rm acc}$. We see that, even for $\beta\,=\,0.6$, $R$ is
less than $r_s$, as our own study assumes. In addition to $\beta$, other 
non-dimensional quantities shown in the table include: $\dot M$, the mass 
accretion rate; ${\dot P}_{\rm grav}$, the gravitational contribution to the 
force; ${\dot P}_{\rm adv}$, the advective component; and their sum 
${\dot P}_{\rm tot}$, which is also the force $F$. Notice that 
${\dot P}_{\rm tot}$ is positive, i.e., it points in the $+z$-direction in 
Figure~\ref{fig:coordinates}. The quantity ${\dot P}_{\rm adv}$ is negative in 
all cases, and smaller in magnitude than ${\dot P}_{\rm tot}$, so that the net
force indeed retards the motion of the mass relative to the background gas.

As the seventh column of the table shows, ${\dot P}_{\rm tot}$ is not equal to 
${\dot M}\,\beta$. Even for \hbox{$\beta\,\sim\,1$}, the ratio
\hbox{${\dot P}_{\rm tot}/{\dot M}\,\beta$} exceeds unity, and inreases with
$\beta$, as found earlier by \citet{s85}. If there is no error in our
theoretical derivation, what could be the source of this discrepancy?

A successful simulation must replicate the flow well in inside of $r_{\rm acc}$.
This task becomes more demanding at higher $\beta$, since $r_{\rm acc}$ itself
varies as $\beta^{-2}$. Following how the gas joins onto the central object is
especially critical. Suppose, for example, that the velocity of material just 
outside the gravitating mass were, for any reason, artificially low in a
simulation. Then, for a given $\dot M$, mass conservation dictates that the
density close to the object be 
increased.
Also increased would be the
value of ${\dot P}_{\rm grav}$, which is obtained by integrating over the 
surrounding density. Conversely, ${\dot P}_{\rm adv}$ would be 
decreased,
since it is proportional to the angle-averaged incoming speed. Both factors 
would cause the numerically determined ${\dot P}_{\rm tot}$ to be too large.

Only a modest lowering of the speed creates a significant rise in 
${\dot P}_{\rm tot}$. To illustrate the point, we correct for this effect by
lowering the numerically calculated ${\dot P}_{\rm grav}$ at each $\beta$-value 
by a factor $f$, where \hbox{$f\,>\,1$}. Simultaneously, we let 
${\dot P}_{\rm adv}$ be increased by the same factor. There is some 
value of $f$ such that ${\dot P}_{\rm tot}$ equals ${\dot M}\,\beta$:
\begin{equation*}
{\dot P}_{\rm grav}/f \,+\, f\,{\dot P}_{\rm adv} \,=\, {\dot M}\,\beta \,\,.
\end{equation*}
This special $f$-value is listed in the last column of Table~1. As $\beta$ 
increases by almost a factor of 20, $f$ is always close to unity and varies 
only from 1.2 to 1.3. Thus, relatively small errors in the two force 
contributions, if they are inversely related, make a big difference in
${\dot P}_{\rm tot}$.

In an earlier paper outlining his numerical method, \citet{r94} stated that he
softened the gravitational potential of the central mass specifically to keep
the incoming velocity relatively low and thereby lengthen the computational 
time step (see his \S 2.4). Needless to say, a softened potential indeed 
creates such an effective deceleration. So does numerical viscosity. In the 
Piecewise Parabolic code that \citet{r94} employed, the viscosity in a region
spanned by $N$ zones scales as $N^{-3}$ \citep[][p.~319]{pw94}. For all his 
isothermal simulations, \citet{r96} used a relatively coarse grid surrounding 
the mass, which typically had five zones covering a distance outside the 
object equal to its radius.  

Since the work of \citet{r96}, other researchers have tackled this flow
problem with improved codes. Results have evolved substantially. From
Table~1, \citet{r96} found \hbox{$\dot M\,=\,0.994$} for \hbox{$\beta\,=\,1.4$}.
In comparison, equation~(\ref{eqn:throop}), which is a fit from \citet{mt09}
to their own simulations, gives \hbox{$\dot M\,=\,0.409$} for the same 
$\beta$-value. A fresh determination of the force, using modern numerical
techniques, would clearly be of interest. 

We stress that both $\dot{M}$ and $\dot{P}_{\rm tot}$ at each $\beta$ should not only be
    calculated in the vicinity of the central mass, but also through a far-field
    integration. Agreement of the results obtained by
these two methods would, of course, strongly corroborate the accuracy of the 
simulation, and also be a sensitive test that the flow has indeed reached 
steady state.\footnote{ Canto et al (2011) ran a simulation for $\beta = 5$ and performed a surface
    integration of the momentum flux over a large sphere. However, they considered
    only the kinetic part of the flux, i.e., the first righthand term in our
    equation (53). The second term, from thermal pressure, is of comparable
    magnitude.}

\subsection{Applications of the New Force Law}

For illustrative purposes, we first consider the deceleration of a gravitating 
mass traveling through a uniform gas, and subject only to the force of
dynamical friction. While the mass moves, it simultaneously accretes gas. The
force is the rate of change of the object's momentum, so that we have the
dimensional relation
\begin{equation*}
\frac{d{\phantom t}}{dt} \left({\dot M}\,V\right) \,=\, - {\dot M}\,V \,\,. 
\end{equation*}
Expansion of the derivative and integration yields
\begin{equation}\label{eqn:vm}
\frac{V}{V_0}\,=\, \left(\frac{M}{M_0}\right)^{\!-2} \,\,,
\end{equation}
where $V_0$ and $M_0$ are, respectively, the object's initial speed and mass.
We previously derived this result in Section~4 of Paper~I.

Pursuing the same reasoning as before, we use $\dot M$ from 
equation~(\ref{eqn:throop}) to derive the non-dimensional equation governing
the velocity evolution:
\begin{equation}
\left(\frac{1}{\beta}\right)\,
\frac{d\beta}{d\tau}  \,=\, -{{4\,\left(\lambda^2\,+\,\beta^2\right)^{1/2}}
\over{\left( 1\,+\,\beta^2\right)^2}} 
\left(\frac{\beta_0}{\beta}\right)^{1/2} \,\,.
\end{equation}
Here $\beta_0$ is the initial speed, and \hbox{$\tau\,\equiv\,t/t_0$} a 
non-dimensional time. We have used as our fiducial dimensional time 
\begin{equation}
t_0 \,\equiv\,\frac{M_0}{2\pi \rho_0 c_s r_s^2} \,\,.
\end{equation}
Here the denominator is the fiducial mass accretion rate from Section~4.4. 
Thus, the quantity $t_0$ is a characteristic growth time for the mass when
\hbox{$\beta \lesssim 1$}, but departs from that time in the supersonic case. 

The upper and lower panels of Figure~\ref{fig:velocity} show the evolution of 
the velocity and mass, respectively, for various $\beta_0$-values. Here we 
obtained the mass using equation~(\ref{eqn:vm}). The figure includes one
subsonic case, \hbox{$\beta_0\,=\,0.5$}. Notice that, when the mass starts out
at supersonic speed, it first decelerates gradually, then much more sharply as
$\beta$ nears unity. Concurrently, its mass climbs rapidly, ultimately 
increasing at the Bondi rate appropriate for a stationary particle.

Turning to more concrete astrophysical applications, the frictional force may
play an important role in the dynamical evolution of planets. In a relatively 
large fraction of exoplanets discovered through radial velocity studies, the 
normal to the orbital plane is misaligned with the stellar spin axis 
\citep{a12}. Recently, \citet{ttp13} have simulated the evolution of an 
inclined, eccentric planet at the early epoch when a circumstellar disk is 
still present. As the planet plunges through the disk periodically, it
experiences an impulsive frictional drag that gradually decreases both its 
orbital inclination and eccentricity, leading eventually to a coplanar, 
circular orbit.

In their numerical study, \citet{ttp13} found that planets with the mass of
Neptune or lower can maintain their orbital inclination over the disk lifetime.
On the other hand, Jovian and higher-mass planets cannot, as a result of the
increased dynamical friction experienced during each disk crossing. The critical
mass separating these two regimes must depend on the prescription for the
frictional drag.

\citet{ttp13} used, for this drag, both that arising from direct impact with 
the gas and the hypersonic limit of the force derived by \citet{o99}. In our 
non-dimensional units, their adopted dynamical friction force $F_T$ is
\begin{equation*}
F_T \,=\, \frac{2\,{\cal I}}{\beta^2} \,\,,
\end{equation*} 
where the factor $\cal I$ is
\begin{equation*}
{\cal I}\,\equiv\, {\rm ln}\,(H/R) \,\,.
\end{equation*} 
Here, $R$ is the radius of the planet and $H$ is the disk's semi-thickness, 
which they used in place of Ostriker's term $V\,t$. The ratio of our derived 
force $F$ to this one is
\begin{equation*}
\frac{F}{F_T} \,=\,\frac{\beta^3 \left(\lambda^2 + \beta^2\right)^{1/2}}
{\left(1 + \beta^2\right)^2\,{\cal I}} \,\,,
\end{equation*}
which approaches ${\cal I}^{-1}$ for \hbox{$\beta \gg 1$}. In their
Section~2.4, \citet{ttp13} quote a characteristic value of 
\hbox{${\cal I}\,\approx\,6$}. Adoption of the new, diminished friction force 
could significantly influence the alignment of orbital planes, and thus the 
fraction of misaligned planets that survive the disk era.

On vastly larger scales, dynamical friction also plays a role in the assembly
of supermassive black holes. Accretion onto these $10^9\,\,\Msun$ objects 
powers the bright quasars detected at redshifts of \hbox{$z\,\sim\,6$}, i.e.,
about 1~Gyr after the Big Bang. The most popular hypothesis is that supermassive
black holes arise through the merger of ``seed'' black holes 
(\hbox{$M\,\sim\,10^2\,\,\Msun$}), which are in turn the remnants of the first
stellar generation \citep{h03}. The mergers occur within the dark matter 
haloes of coalescing galaxies. An individual black hole also accretes gas that
has settled toward the center of its parent halo.

In a combined semi-analytic and numerical study, \citet{th09} showed that seed
black holes can indeed form a supermassive one in the requsite time, provided 
they remain embedded in the halo's gas component. Mass buildup is delayed 
by the large recoil velocities in the coalesced product of each black hole 
merger. (Asymmetric gravitation radiation carries off the remaining momentum.)
Promoting the merger process is dynamical friction, which can return far-flung 
black holes to the halo center. For their detailed numerical simulations, 
\citet{th09} assumed the total friction force to be
\begin{equation*}
F_{\rm tot} \,=\, F_{\rm DF} \,+\, {\dot M}\,V \,\,,
\end{equation*}
where $\dot M$ is the mass accretion rate of surrounding gas. They took the 
dynamical friction force $F_{\rm DF}$ to be the sum of that due to the 
collisionless sea of dark matter and that created by the gas. For the latter, 
\citet{th09} adopted the formula of \citet{o99}. 

According to our own study, the last term above encompasses the entire gaseous
dynamical friction. That is, the dynamical friction force assumed in the 
simulations is too large in magnitude. A diminished $F_{\rm tot}$ would increase 
the frequency of black hole ejections, as well as the time over which black 
holes flung outward remain on their extended orbits. In summary, the growth of 
supermassive black holes is delayed. As in the planetary problem, a revised 
calculation incorporating the corrected dynamical friction would be of interest.

\section{Discussion}

Two key assumptions underlie our derivation, and we reiterate these as a
cautionary note. The first is that the geometric size of the gravitating mass
be appropriately small. Specifically, we assume that $R \ll r_{\rm acc}$. In the opposite extreme, $R \gg r_{\rm acc}$, 
    the main force acting on the object is not gravitational in origin, but comes 
    from the direct impact of surrounding gas. This impact force is given dimensionally
    by $F_{\rm imp} \approx \pi\rho_0R^2V^2$. The ratio of this force to that of dynamical friction is
	\begin{equation}
		\frac{F_{\rm imp}}{F} \approx  \frac{1}{\dot{M}\beta^3} \left(\frac{R}{r_{\rm acc}}\right)^2\ ,
	\end{equation}
    where $\dot{M}$ is non-dimensional. According to equation (51), the product $\dot{M}\beta^3$
    approaches a constant for $\beta \gg 1$.
    
    Our second assumption is that the gas surrounding the object be 
sufficiently rarefied that its self-gravity is negligible. Once the amount of
mass within the radius $r_{\rm acc}$ becomes comparable to or exceeds that of the 
central object, the problem becomes one of gravitational collapse. Here,
accretion generally does not occur in a steady-state manner 
\citep[see also Section~5 of][]{l13}.  

The present dynamical friction calculation differs sharply both in approach
and result from previous efforts. Starting with \citet{bh44}, all other
researchers have obtained expressions for the force that contain a Coulomb
logarithm. This term arises from integrating over the perturbed gas, usually
in the downstream wake \citep{bh44,o99}. \citet{d64} also obtained this term,
but by integrating the acoustic energy flux over a large sphere surrounding
the gravitating mass. His derivation is thus closer in spirit to ours. 
However, we are dubious of his basic premise that the total energy loss 
associated with the moving mass is fully accounted for by this outgoing, 
acoustic disturbance. Accretion generates shocks close to the mass, and the 
power from radiating shocks could in principle rival the acoustic loss, for a 
sufficiently high velocity. On the other hand, the total inflow of linear 
momentum is always transported unaltered to the object itself, provided the 
surrounding flow is steady-state.

The strategy we adopted was to use a spherical coordinate system and
expand the far-field perturbations of the stream function and density
in power series in the radius $r$.  We found that various coefficients
in these series diverged at both the Mach and anti-Mach cones. Since
the series themselves are only valid far from the gravitating mass,
these divergences are purely mathematical, and result from the
specific form of the perturbation series. It may be, therefore, that
all divergences could be eliminated through an appropriate change of
independent variables. We leave this technical issue as a topic of
future study.

Turning to our specific result, we find it compelling that the friction force
in the present, supersonic case is identical in form to that found in Paper~I
for objects moving subsonically. Our final expression is, in fact, so simple
that it raises the question of whether the complex machinery we brought to
bear was necessary for its derivation. As one alternative path, consider the
fact that the velocity $V$ is the momentum per unit mass carried by the
background gas. Multiplying $V$ by $\dot M$, the mass accreted per time, 
immediately yields the rate of momentum input, which is the force.

This concise derivation is intuitively appealing, but is also incorrect.
While $V$ is the 
limiting
velocity of background gas at infinite 
distance from the mass, the true velocity differs at any finite $r$. Similarly,
the true density differs from its background value. These differences are
relatively small, but they are integrated over a large surface to obtain the 
net inflow of momentum. As may be seen in Section~5.1, this inflow is 
${\dot M}\,V$  plus several correction terms of comparable magnitude. These 
additional terms cancel exactly, by virtue of the fact that the flow is 
irrotational. A more explicit demonstration of the analogous, subsonic result 
is in Section~5.2 of Paper~I. 

In any event, we do agree that a simpler derivation of our result should exist,
provided the far-field flow is irrotational. Equivalently, the generalized
derivation must apply to barytropic fluids, i.e., those in which $P$ is a
function only of $\rho$. In support of this contention, we note that
\citet{kd12} utilized our technique of far-field integration to show that
\hbox{$F\,=\,{\dot M}\,V$} holds for an object moving subsonically through an
isentropic fluid with arbitrary adiabatic index $\gamma$. We expect that 
analogous reasoning will yield the same result for an isentropic fluid in the 
supersonic regime. At that point, the stage will be set for the new derivation,
one that avoids a detailed description of the far-field flow. Whether one will 
still need to assume a steady-state mass accretion rate, as we have done, or be
able to derive this rate from first principles, remains to be seen.

\begin{acknowledgements}
We thank Max Ruffert for a thoughtful referee report that helped improved the clarity of the paper. 
Throughout this project, ATL was supported by an NSF Graduate Fellowship, while 
SWS received partial funding from NSF Grant~0908573.
\end{acknowledgements}

\clearpage

\begin{table}
\caption{Results of \citet{r96}}
\begin{tabular}{ccccccccc}
\hline\hline
$R$ & $\beta$ & $\dot{M}$ & $\dot{P}_{\rm grav}$ & $\dot{P}_{\rm adv}$ 
& $\dot{P}_{\rm tot}$ & $\dot{P}_{\rm tot}/\dot{M}\beta$
& $f$ \\
\hline
1.11(-1)   &   0.6	  & 2.41  & 3.19  
& -9.82(-1) & 2.21  & 1.53  & 1.21 \\
2.04(-1)   &   1.4	& 9.94(-1) & 1.08(+1)
& -6.63 & 4.14 & 2.98	 & 1.17 \\
4.44(-3) & 	3.0	& 7.86(-2) & 2.20  & -1.26
& 9.43(-1)  & 4.00	 & 1.23 \\
4.00(-4) &	10.0&	1.59(-3)  & 1.78(-1) 
& -8.91(-2) & 8.91(-2) & 5.60 & 1.33 \\
\hline
\end{tabular}
\tablefoot{
For the definitions of the quantities displayed, see Section 5.2 of the text. 
All entries are non-dimensional. Figures in parentheses are the power of ten multiplying the preceding number.}
\label{tab:Ruffert}
\end{table}

\clearpage

\appendix

\section{Background and First-Order Contributions to the Momentum Influx}

Here we evaluate ${\dot P}_2$ and ${\dot P}_1$, the first two coefficients on
the righthand side of equation~(\ref{eqn:Fseries}). We will show that both
coefficients are zero. The first is
\begin{equation}
{\dot P}_2 \,=\, -\int_0^\pi 
\left[\left(f_2^\prime\right)^2 {\rm cot}\,\theta \,+\,
2\,f_2\,f_2^\prime \,+\,
{\rm cos}\,\theta\,\,{\rm sin}\,\theta\right] d\theta \,\,.
\end{equation}
Substituting \hbox{$f_2\,=\,\beta\,{\rm sin}\,\theta /2$}, this becomes
\begin{eqnarray*}
{\dot P}_2 \,&=&\,- \left(1\,+\,\beta^2\right)
\int_0^\pi\!{\rm sin}\,\theta\,\,{\rm cos}\,\theta\,\,d\theta \\
&=&\,0  \,\,.
\end{eqnarray*}
The background flow, while it certainly carries $z$-momentum, 
makes no 
net
contribution to the influx.

The integral expression for the second coefficient is lengthier, and we write 
it as
\begin{equation}\label{eqn:Pdot1}
{\dot P}_1\,=\,-\int_0^\pi \left({\cal C}_1 \,+\,{\cal C}_2\right)
\,d\theta \,\,,
\end{equation}
where
\begin{equation*}
{\cal C}_1 \,\equiv\,2\,f_2^\prime\,f_1^\prime\,{\rm cot}\,\theta \,\,+\,\,
2\,f_2\,f_1^\prime  \,\,+\,\,f_2^\prime\,f_1 \,\,,
\end{equation*}
and
\begin{equation*}
{\cal C}_2 \equiv 
g_{-1}\,{\rm sin}\,\theta\,\,{\rm cos}\,\theta \,\,-\,\,
\left(f_2^\prime\right)^2 g_{-1}\,{\rm cot}\,\theta \,\,-\,\,
2\,f_2\,f_2^\prime\,\,g_{-1} \,\,. 
\end{equation*}

It is convenient to split up the integral in equation~(\ref{eqn:Pdot1}) by 
region. For the downstream part, which we denote as ${\dot P}_1^d$, we use
equations~(\ref{eqn:fFOup}) and (\ref{eqn:gFOup}) for $f_1$ and $g_{-1}$. We
find, after algebraic simplification, that
\begin{equation}
{\dot P}_1^d \,=\, -\int_0^{\theta_M}\! d\theta
\left(\frac{\beta^2\,{\rm sin}\,\theta\,{\rm cos}\,\theta}
{{\cal D}^{1/2}} \,+\, {\rm cos}\,\theta\,\,{\rm sin}\,\theta\right) \,\,.
\end{equation}
We now use the identity
\begin{equation*}
\frac{d\,{\cal D}^{1/2}}{d\theta} \,=\, 
-\frac{\beta^2\,{\rm sin}\,\theta\,\,{\rm cos}\,\theta}{{\cal D}^{1/2}} \,\,,
\end{equation*}
to conclude that
\begin{align}
\begin{split}
{\dot P}_1^d \,&=\,\int_0^{\theta_M}\!\! d\,{\cal D}^{1/2} \,-\,
\int_0^{\theta_M}\!\! d \left({\rm sin}^2\,\theta/2\right) \\
&=\, -1 \,-\, \frac{1}{2\,\beta^2} \,\,.
\end{split}
\end{align}

Similarly, the upstream part is
\begin{align}
\begin{split}
{\dot P}_1^u \,&=\,\int_{\theta_M^\prime}^\pi\!\! d\,{\cal D}^{1/2} \,-\,
\int_{\theta_M^\prime}^\pi\!\! d \left({\rm sin}^2\,\theta/2\right) \\
&=\, 1 \,+\, \frac{1}{2\,\beta^2} \,\,,
\end{split}
\end{align}
so that
\begin{equation}
{\dot P}_1^d \,+\,  {\dot P}_1^u \,=\, 0 \,\,.
\end{equation}

For the intermediate piece, to be denoted ${\dot P}_1^i$, we use 
equations~(\ref{eqn:fFOiC2}) and (\ref{eqn:gFOiC2}) for $f_1$ and $g_{-1}$.
Although we have established that \hbox{$C\,=\,+1$}, we will see that
${\dot P}_1^i$ is independent of this parameter, so we retain the original
notation. After algebraic manipulation, we find
\begin{equation}
{\dot P}_1^i \,=\, -\int_{\theta_M}^{\theta_M^\prime}\!d\theta \left( 
\frac{\beta\,C\,{\rm sin}\,\theta\,{\rm cos}\,\theta}{{\cal E}^{1/2}} \,+\,
{\rm cos}\,\theta\,\,{\rm sin}\,\theta \right) \,\,.
\end{equation} 
Using the identity
\begin{equation*}
\frac{d\,{\cal E}^{1/2}}{d\theta} \,=\, 
\frac{\beta^2\,{\rm sin}\,\theta\,\,{\rm cos}\,\theta}{{\cal E}^{1/2}} \,\,,
\end{equation*}
we infer that
\begin{align}
\begin{split}
{\dot P}_1^i \,&=\, -C \int_{\theta_M}^{\theta_M^\prime}\!d{\cal E}^{1/2} \,-\,
\int_{\theta_M}^{\theta_M^\prime}\!d\left({\rm sin}^2\,\theta /2\right) \\
&=\, 0 \,\,. 
\end{split}
\end{align}
In summary, we have
\begin{equation}
{\dot P}_1 \,=\, {\dot P}_1^d \,+\, {\dot P}_1^u \,+\, {\dot P}_1^i \,=\, 0
\,\,. 
\end{equation}

\newpage

\begin{figure}
\includegraphics[width=\textwidth/2]{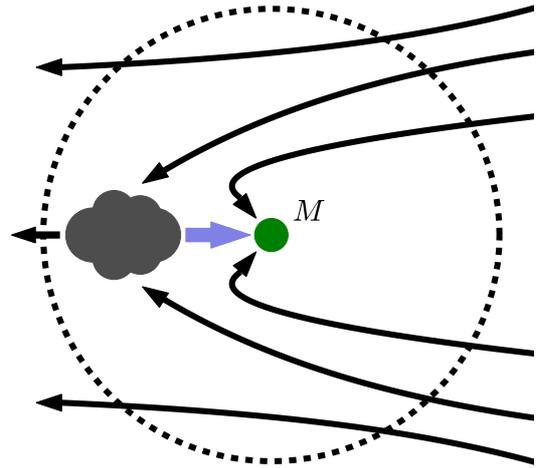}
\caption{Sketch of mass and momentum flow. Surrounding the central gravitating body of mass $M$ is a large, imaginary sphere. The thin curves represent streamlines of background gas. After entering the sphere, most gas simply exits again, while a small portion joins directly onto the mass. In addition, some gas temporarily forms an overdense wake, also sketched here. The wake tugs gravitationally on the mass, thereby imparting momentum to it (broad arrow). All gas entering the wake also leaves it, either joining the mass or exiting the sphere.}
\label{fig:cartoon} 
\end{figure}

\begin{figure}
\includegraphics[width=\textwidth/2]{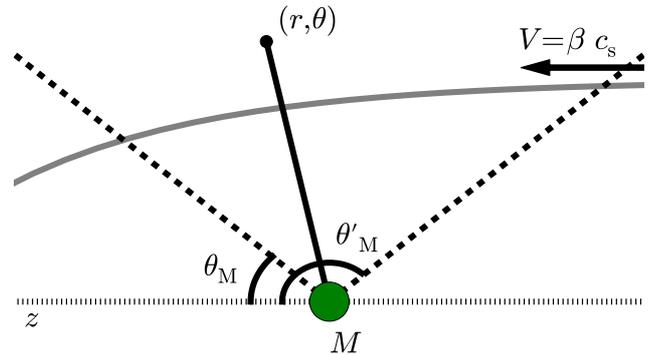}
\caption{Mathematical treatment of the flow. We erect a spherical coordinate 
system centered on the gravitating body. The gas is isothermal, and 
its velocity far upstream is $\b\,\cs\ (\b>1)$. Indicated are the Mach angle
$\theta_M$ and its supplement, 
\hbox{$\theta_M^\prime \,\equiv\,\pi\,-\,\theta_M$}.}
\label{fig:coordinates} 
\end{figure}

\begin{figure}
\includegraphics[width=\textwidth/2]{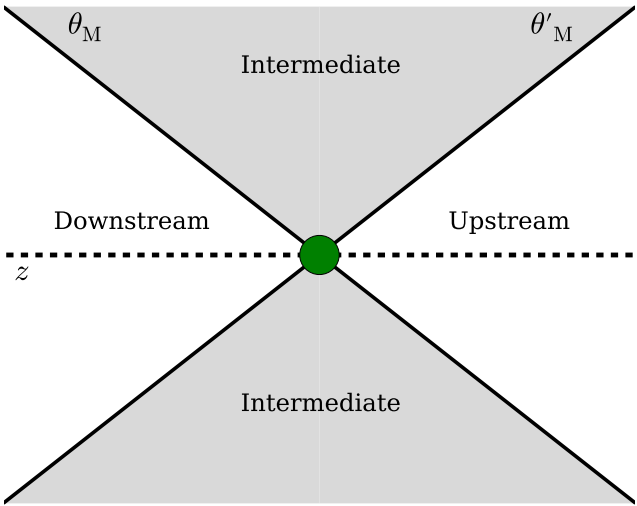}
\caption{Regions of the flow. The upstream and downstream regions lie 
within the anti-Mach and Mach cones, respectively. Between the two cones is
the intermediate region. In the far-field flow, there are no
physical barriers between the intermediate region and its neighbors,
although mathematical divergences appear at the two cones.}
\label{fig:regions}
\end{figure}

\begin{figure}
\includegraphics[width=\textwidth/2]{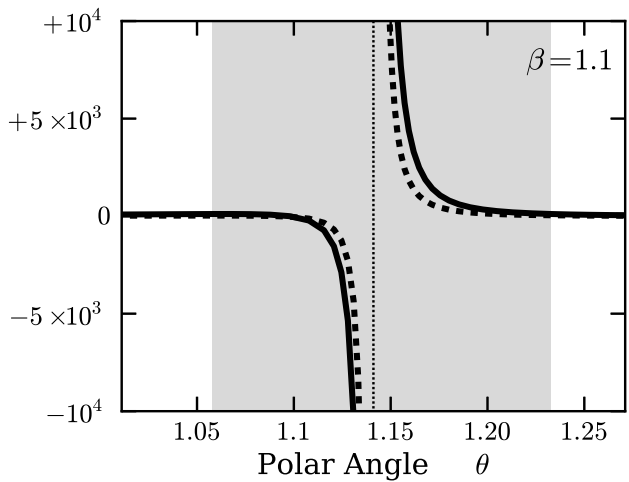}
\caption{Asymmetric divergences at the Mach angle, for \hbox{$\beta\,=\,1.1$}.
The solid and dashed curves trace $f_0^\prime$ and $g_{-2}$, respectively, close
to the Mach angle $\theta_M$, marked here by the vertical, dotted line. The
functions diverge antisymmetrically once we choose \hbox{$C^2\,=\,1$}. Both
curves are taken from the numerical integration outlined in Section~4.5. The
shaded region is that which we remove before enforcing continuity of $f_0$ 
and $g_{-2}$ across the cones.}  
\label{fig:diverge}
\end{figure}

\begin{figure}
\includegraphics[width=\textwidth/2]{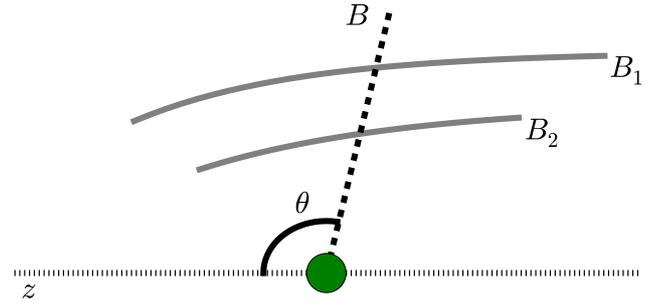}
\caption{Enforcing irrotationality. In our steady-state, isothermal flow, the
Bernoulli function $B$ is constant along each streamline. In principle, $B$
could vary from one streamline to the next. However, if we also make $B$
constant along any cone, then the function is a universal constant and the
flow is irrotational.}
\label{fig:vort}
\end{figure}

\begin{figure}
\includegraphics[width=\textwidth/2]{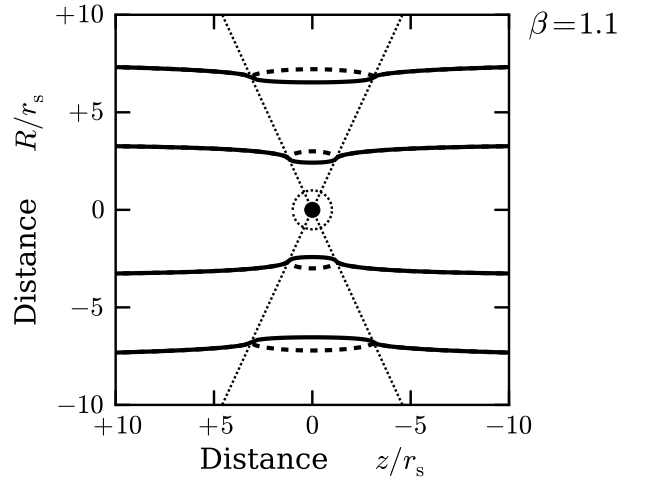}
\caption{First-order streamlines. Shown are contours of constant
\hbox{$f_2 \,r^2\,+\,f_1\,r$}, for the case \hbox{$\beta\,=\,1.1$}. The dotted,
diagonal lines trace the Mach and anti-Mach cones, while the central, dotted
circle is \hbox{$r/r_s\,=\,1$}. Within the intermediate region, the dashed
streamlines were constructed assuming \hbox{$C\,=\,-1$}. The solid ones 
correspond to \hbox{$C\,=\,+1$} and are more realistic.}
\label{fig:fos11}
\end{figure}

\begin{figure}
\includegraphics[width=\textwidth/2]{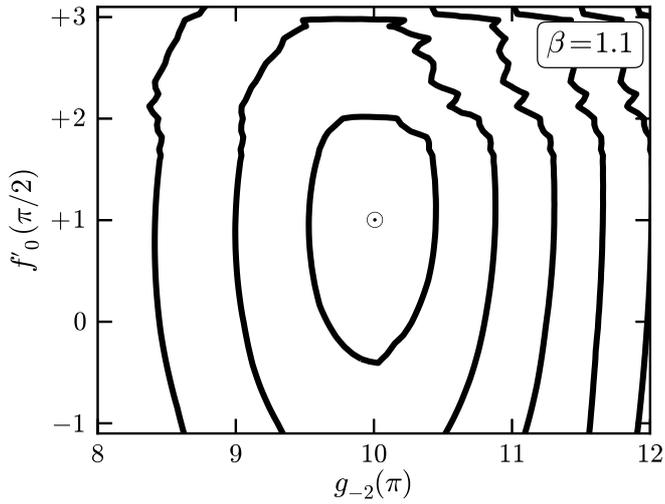}
\caption{Contours of constant $\chi$, where $\chi$ is defined by
equation~(\ref{eqn:chi}) in the text. Adjacent contours are separated by
a $\chi$-interval of 15.0. The minimum point, indicated by the central dot 
within the circle, is \hbox{$f_0^\prime (\pi/2)\,=\,0.96$} and 
\hbox{$g_{-2} (\pi)\,=\,10.0$}. The $\chi$-value of this point is 2.22.}
\label{fig:contour}
\end{figure}

\begin{figure}
\includegraphics[width=\textwidth/2]{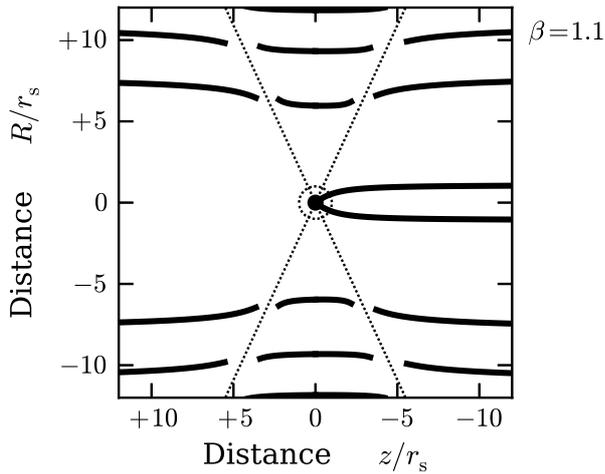}
\caption{Streamlines of the second-order flow, for \hbox{$\beta\,=\,1.1$},
constructed using the best-fit parameters from Figure~\ref{fig:contour}. The 
shaded regions surrounding the two Mach cones represent sectors in which the 
perturbation coefficients diverge. Also indicated are the central mass and the 
surrounding circle corresponding to \hbox{$r\,=\,1$}. Notice that the innermost
streamlines reach the origin, indicating the occurrence of mass accretion.}
\label{fig:sos11}
\end{figure}

\begin{figure}
\includegraphics[width=\textwidth/2]{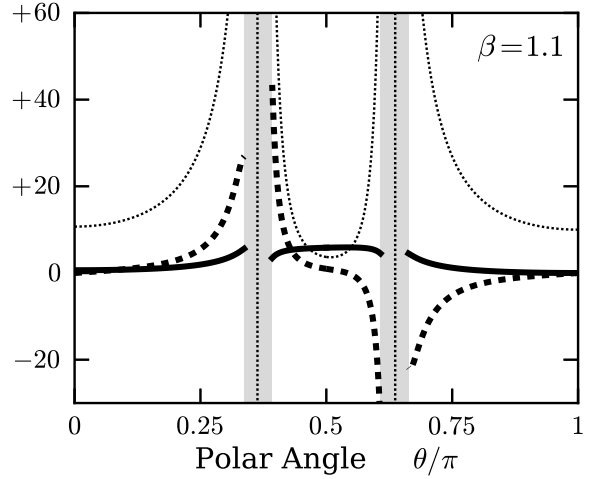}
\caption{Angular variation of perturbation coeffficients for 
\hbox{$\beta\,=\,1.1$}. The dashed curves show $f_0^\prime$, while the dotted
curves are $g_{-2}$, again using the best-fit parameters from 
Figure~\ref{fig:contour}. Finally, the solid curve is the integrated $f_0$. As 
in Figure~\ref{fig:sos11}, all variables diverge within the shaded regions
straddling the two Mach cones.}
\label{fig:gfp11}
\end{figure}

\begin{figure}
\includegraphics[width=\textwidth/2]{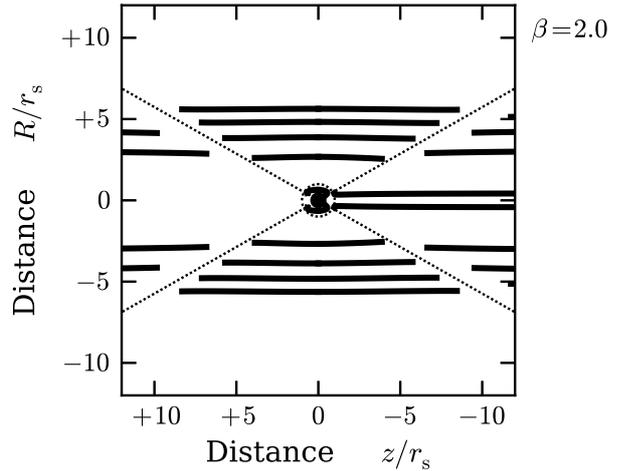}
\caption{Streamlines of the second-order flow for \hbox{$\beta\,=\,2.0$},
constructed using \hbox{$f_0^\prime (\pi/2)\,=\,0.06$} and
\hbox{$g_{-2} (\pi)\,=\,-0.54$}. In this case, fluid elements are nearly 
following linear trajectories. The innermost streamlines again reach the 
central mass at the origin.}
\label{fig:sos20}
\end{figure}

\begin{figure}
\includegraphics[width=\textwidth/2]{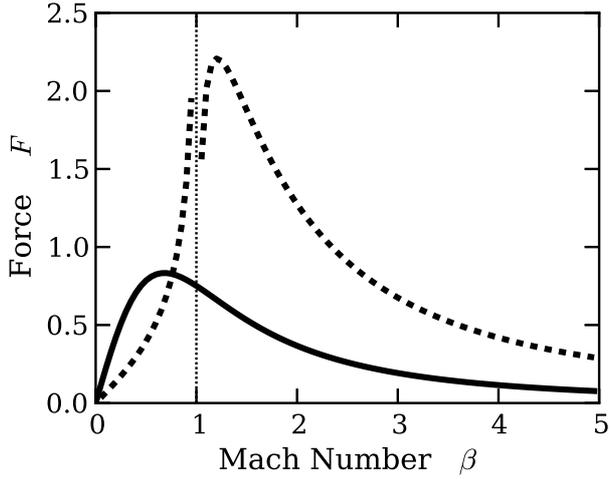}
\caption{The dynamical friction force $F$, shown as a function of the Mach
number $\beta$. Our calculated force, taken from equation~(\ref{eqn:Fthroop}),
incorporates the prescription for $\dot M$ from \citet{mt09}. The dotted 
vertical line marks \hbox{$\beta \,=\,1$}. Also shown by the broken dashed 
curve is the force as calculated by \citet{o99}. This force diverges at 
\hbox{$\beta\,=\,1$}.} 
\label{fig:force}
\end{figure}

\begin{figure}
\includegraphics[width=\textwidth/2]{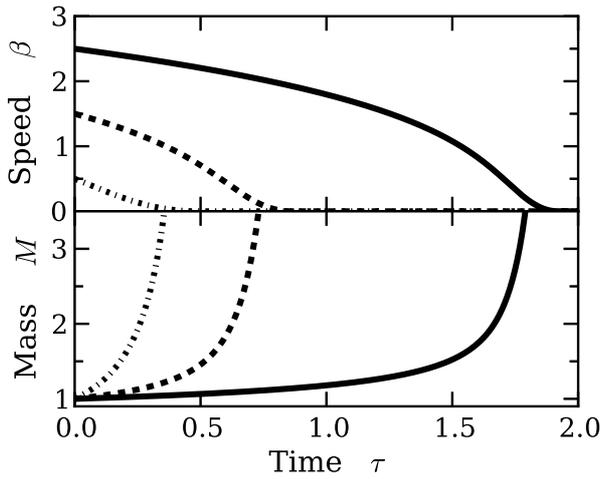}
\caption{Evolution of the particle's speed and mass as a function of
non-dimensional time $\tau$. Mass is shown relative to its initial value. The
different curves represent different initial speeds: $\beta_0\,=\,0.5$, 1.5, 
and 2.5. Once the speed of an initially supersonic particle approaches 
\hbox{$\beta\,\sim\,1$}, it quickly decelerates. Concurrently, its mass grows 
rapidly.}
\label{fig:velocity}
\end{figure}

\clearpage

\end{document}